\documentclass[aps,prb,showpacs,preprintnumbers,amsmath,amssymb]{revtex4}
\usepackage[dvips]{graphicx}
\usepackage{dcolumn}
\usepackage{bm}

\begin{document}

\title{
Possible Triplet Superconducting Order in Magnetic Superconducting Phase induced by Paramagnetic Pair-Breaking 
}

\author{Ken-ichi Hosoya}
\author{Ryusuke Ikeda}

\affiliation{
Department of Physics, Kyoto University, Kyoto 606-8502, Japan
}

\date{\today}

\begin{abstract}
Motivated by recent thermal conductivity measurements in the superconductor CeCoIn$_5$, we theoretically examine a possible staggered spin-triplet superconducting order to be induced by the coupled spin-density-wave (SDW) and $d$-wave superconducting (SC) orders in the high field and low temperature (HFLT) SC phase peculiar to this material with strong paramagnetic pair-breaking (PPB). It is shown that one type of the $\pi$-triplet order is consistent with that explaining the thermal conductivity data and can naturally be incorporated in the picture that the $Q$-phase is a consequence of the strong PPB effect inducing the SDW order and the FFLO spatial modulation parallel to the applied magnetic field. 
\end{abstract}

\maketitle

\section{Introduction}
\label{sec:introduction}

The high field and low temperature (HFLT) superconducting (SC) phase \cite{Bianchi}, the so-called $Q$-phase, of the $d_{x^2-y^2}$-wave paired superconductor CeCoIn$_5$ continues to show strange phenomena, and its nature is still a matter under much debate. Data of NMR measurements \cite{Kumagai1} and the doping experiment \cite{Tokiwa} have shown results consistent with the presence in this phase of the amplitude of the SC order parameter modulated spatially {\it along} the magnetic field \cite{RIfragile}. On the other hand, it is known \cite{Kenzel} that a long range spin density wave (SDW) order with a ${\bm Q}$-vector parallel to a gap node of the $d_{x^2-y^2}$-wave pairing function is present in the HFLT phase and disappears as the SC order is lost by increasing the field. It is natural to expect the strong paramagnetic pair-breaking (PPB) effect seen clearly in, e.g., the $H_{c2}(T)$ curve \cite{SIKEDA} and the discontinuous nature of the mean field $H_{c2}$-transition at lower temperatures \cite{Izawa}, in this material is the main origin of such strange properties. In fact, it is plausible that the suggested \cite{RIfragile} spatial modulation of the SC order is attributed to the presence of the PPB-induced Fulde-Ferrell-Larkin-Ovchinnikov (FFLO) SC order in the HFLT phase \cite{Ada,Yanase}. Further, the presence of a basic mechanism inducing the SDW order \cite{Kenzel,Kumagai1} based on the strong PPB in the $d$-wave paired SC phase has been noticed \cite{IHA}. It has been stressed in Ref.10 that, although this PPB-induced SDW ordering is essentially of an electronic origin \cite{other,Hatakefinal}, it is enhanced by the FFLO spatial modulation of the amplitude of the SC order parameter. 

A different picture on the SDW order in the HFLT phase is based not on the presence of the strong PPB but on the assumption of a $\pi$-triplet SC order present only in such high fields \cite{Agter,Aperis}. Though this approach has been used as the simplest picture explaining the original neutron scattering measurements \cite{Kenzel}, the assumption that the $\pi$-triplet order inducing the SDW order spontaneously occurs in such higher fields lacks a concrete support based on a reasonable microscopic model and has not been justified so far. In fact, the recent detection that the SDW ${\bm Q}$-vector favors the nodal direction perpendicular to the field \cite{Gerber} is found not to be explained based on this scenario. Rather, this ${\bm Q}$-vector orientation sensitive to the field direction has been microscopically explained as a pinning effect of the ${\bm Q}$ vector to the FFLO nodal planes perpendicular to the field \cite{Hatake2015}. 

However, the recent thermal conductivity data have shown a feature which cannot be explained without the $\pi$-triplet SC order in the HFLT phase \cite{Kim}. Upon rotating the magnetic field direction through the [100]-direction within the basal plane, the thermal conductivity jumps together with the discontinuous change of the SDW ${\bm Q}$-vector \cite{Gerber} when ${\bm H} \parallel$ [100]. This suggests the presence of an additional SC gap node determined by the SDW ${\bm Q}$-vector. A possible approach will be to extend the theoretical picture \cite{IHA} based on the strong PPB to the case with a $\pi$-triplet SC order. This is not a formidable task because the coexistence of the $d$-wave SC order and a SDW order with ${\bm Q}$-vector parallel to a $d$-wave SC gap node can induce a $\pi$-triplet SC order. 

In the present work, we investigate a possible $\pi$-triplet SC order and its roles in the HFLT phase of CeCoIn$_5$ within the theoretical approach \cite{IHA} based on the strong PPB. 
We find that the $\pi$-triplet order determined theoretically is consistent with that suggested from the thermal conductivity result \cite{Kim}. It is pointed out that the PPB-based theoretical picture on the HFLT phase constructed previously \cite{IHA,Hatake2015} is not changed essentially by taking account of this triplet order, and that inclusion of the $\pi$-triplet order improves the results on the phase diagram in previous works \cite{Hosoya,Hatake1} in a couple 
of ways. The picture obtained in Refs.10 and 16 and its extension done in the present work is summarized in Fig.1. 

\begin{figure}[htb]
 \begin{minipage}{\hsize}
  \begin{center}
   \includegraphics[width=80mm]{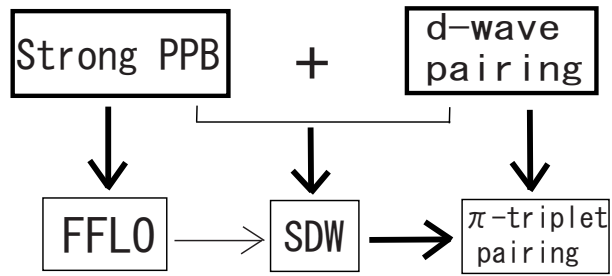}
\end{center}
 \end{minipage}
 \caption{Relations among the PPB-induced multiple orderings to occur in $d_{x^2-y^2}$-wave paired superconductors. The thin arrow indicates such an indirect effect \cite{IHA} that the FFLO spatial modulation assists the SDW ordering, while a direct effect from one event to another is indicated by each thick solid arrow. The present SDW ordering occurs as a combined effect of the PPB and the $d_{x^2-y^2}$-wave pairing. 
}
  \label{fig.1}
\end{figure}

This paper is organized as follows. 
In sec.II, possible staggered $\pi$-triplet SC orders which may occur in the HFLT phase of CeCoIn$_5$ are classified based on the group theoretical method. 
In sec.III, a stable $\pi$-triplet order is examined within the mean field approximation neglecting the FFLO spatial modulation. In sec.IV, the switching of the SDW ${\bm Q}$-vector upon rotation of the in-plane magnetic field is explained in the FFLO theory neglecting the presence of the 
vortices. In sec.IV, effects of the $\pi$-triplet order on the HFLT phase composed of the SDW and FFLO orders are investigated. Further, a summary of the present work is mentioned in sec.V, and details of calculation in sec.III are presented in Appendix. 

\section{Possible triplet order}

First, let us start our analysis from classfying candidates of the $\pi$-triplet orders based on the group theory. Our treatment closely follows the approach by Agterberg et al.\cite{Agter}. In the present context, three order parameters can be realized in the HFLT phase of CeCoIn$_5$. These are the $d$-wave SC order parameter 
\begin{equation}
{\it \Delta} = \frac{|g|}{2} \sum_{\alpha, \beta, {\bm p}} (-i \sigma_y)_{\alpha, \beta} \psi({\bm p}) \langle c_{{\bm p},\alpha} c_{-{\bm p}, \beta} \rangle
\end{equation}
with a scalar pairing function $\psi({\bm p})$, the SDW order parameter 
\begin{equation}
m = \frac{U}{2} \sum_{\alpha, \beta, {\bm p}} ({\bm{\sigma}}\cdot{\hat {\bm n}})_{\alpha, \beta} \langle c^\dagger_{{\bm p},\alpha} c_{{\bm p}+{\bm Q}, \beta} \rangle 
\end{equation}
with the polarization direction ${\hat {\bm n}}$ of the SDW moment, and the staggered $\pi$-triplet SC order parameter 
\begin{equation}
{\tilde {\it \Pi}}^{(s)}_{\pm{\bm Q}} = \frac{V_s}{2} \sum_{\alpha, \beta, {\bm p}} {\bm d}_{s}({\bm p}) \cdot (-i \bm{\sigma} \sigma_y)_{\alpha, \beta} \langle c_{{\bm p} \pm {\bm Q}/2,\alpha} c_{-{\bm p} \pm {\bm Q}/2, \beta} \rangle,
\end{equation}
where the index $s$ indicates the type of the possible $\pi$-triplet order (see below). 

In this section, possible spatial modulations of ${\it \Delta}$ with long wavelengths are neglected for simplicity because they play no essential role for determining a pairing symmetry. In fact, in experiments on CeCoIn$_5$, the pattern of the vortex lattice modulation in the plane perpendicular to the magnetic field is not changed upon entering the HFLT phase by increasing the field \cite{RE}, indicating that some phenomena in the HFLT phase may be described by neglecting the orbital pair-breaking effect of the magnetic field. 

The in-plane component of the SDW ${\bm Q}$ vector will be assumed hereafter to be either of $(k, \pm k)$. The SDW ${\bm Q}$-vector is the sum of the commensurate component ${\bm Q}_0$ and the incommensurate part ${\bm q}$ which is parallel \cite{Kenzel,Hatake1} to ${\bm Q}_0$, and the in-plane component of ${\bm Q}_0$ is either of $(\pi, \pm \pi)$. 

We have the following two possibilities of a third order coupling term in the free energy among the three order parameter fields, 
\begin{equation}
f_1 = -i \, 
m({\it \Delta}^* {\tilde {\it \Pi}}^{(s)}_{-{\bm Q}} - {\it \Delta} ({\tilde {\it \Pi}}^{(s)}_{\bm Q})^*) + {\rm c.c.},
\end{equation}
and
\begin{equation}
f_2 = H m ({\it \Delta}^* {\tilde {\it \Pi}}^{(s)}_{-{\bm Q}} + {\it \Delta} ({\tilde {\it \Pi}}^{(s)}_{\bm Q})^*) + {\rm c.c.},
\end{equation}
where $H$ is the magnitude of the applied magnetic field. 
Through one of $f_j$ ($j=1$, $2$), one order is {\it induced} by the presence of the remaining two orders. Agterberg et al.\cite{Agter,Aperis} have assumed a nonvanishing $\pi$-triplet order as the primary order in the HFLT phase and a nonzero SDW order as the secondary one induced by the primary one. In the present work, the origin of the nonvanishing SDW order is assumed to consist in the strong PPB according to the previous work \cite{IHA}, and a $\pi$-triplet order induced by such a nonvanishing SDW order is taken to be the secondary one 
(see Fig.1). 

\begin{table}[htb]
\begin{center}
\caption{List of irreducible representations and their basis functions in the space group P4/mmm with ${\bm Q}=(k,\pm k,0.5)$ or $(k,\pm k,-0.5)$}
\small
\begin{tabular}{cccccccc} \hline
    Irreducible rep. & ${D}_{\Gamma_s}(E)$ & ${D}_{\Gamma_s}(\sigma_{z})$ & ${D}_{\Gamma_s}(C_{2\eta})$ & ${D}_{\Gamma_s}(\sigma_{\zeta})$ & $\psi({\bm p})$ & ${\bm d}_{s}({\bm p})$ &  ${\bm S}$-component \\ \hline
    $\Gamma_1$  & $1$ & $1$ & $1$ & $1$ & $s, p_{x}p_{y}$ & $\hat{\bm z}(p_{x} \mp p_{y}), p_{z}(\hat{\bm x} \mp \hat{\bm y})$ & \\
    $\Gamma_2$  & $1$ & $1$ & $-1$ & $-1$ & $p_{x}^{2}-p_{y}^{2}$ & $\hat{\bm z}(p_{x} \pm p_{y}), p_{z}(\hat{\bm x} \pm \hat{\bm y})$ & $S_{z}$ \\
    $\Gamma_3$  & $1$ & $-1$ & $-1$ & $1$ & $p_{z}(p_{x} \mp p_{y})$ & $\hat{\bm x}p_{x}-\hat{\bm y}p_{y}, \hat{\bm x}p_{y}-\hat{\bm y}p_{x}$ & $S_{x} \mp S_{y}$ \\
    $\Gamma_4$  & $1$ & $-1$ & $1$ & $-1$ & $p_{z}(p_{x} \mp p_{y})$ & $\hat{\bm x}p_{x}+\hat{\bm y}p_{y}, \hat{\bm x}p_{y}+\hat{\bm y}p_{x}$ & $S_{x} \pm S_{y}$ \\ \hline
  \end{tabular}
\label{Irrep(1-10)}
  \end{center}
\end{table}

Next, the order parameters will be classified in the group-theoretical manner \cite{Agter}. The full space group of CeCoIn$_5$ is P4/mmm. For a given SDW ${\bm Q}$, the two pairing functions, the scalar $\psi$ and the vector ${\bm d}_s$, are defined together with a magnetic vector field ${\bm S}$ as the irreducible representations of the set of four operations conserving ${\bm Q}$. Both the magnetic field and the SDW moment are regarded as one of ${\bm S}$ in this classification. When ${\bm Q}=(k, \pm k, 0.5)$, the four operations including the identity consist of the $\pi$-rotation $C^{(\pm)}_{2 \eta}$ around the axis $(1,\pm 1,0)$, the mirror operation $\sigma_z$ at the basal plane, and the mirror operation $\sigma_{\pm \zeta}$ at the plane perpendicular to $(1,\mp 1,0)$. Here, we only have to extend Table I in the previous work\cite{Agter} to the manner including the case with ${\bm Q}=(k,-k,0.5)$. The resulting Table for ${\bm Q}=(k,\pm k, 0.5)$ is given in Table I, where the $z$-direction in the spin space is taken to be the $c$-axis. 

Based on this Table, the set of the order parameter fields making the coupling term $f_1$ nonvanishing will be first determined. It is known that the SDW moment parallel to the $c$-axis and the $d_{x^2-y^2}$-wave singlet pairing with $\psi({\bm p}) \propto p_x^2 - p_y^2$ are realized in the HFLT phase of CeCoIn$_5$. Thus, the only staggered $\pi$-triplet pairing leading to a nonvanishing coupling term (4) belongs to $\Gamma_1$ and, when ${\bm Q} = (k, \pm k,0.5)$, is given in the representation (3) by ${\bm d}_1 \propto {\hat {\bm z}} (p_x \mp p_y)$. Namely, the gap node of the ${\bm d}_1$-vector is always directed to the SDW ${\bm Q}$-vector. Note that the two ${\bm d}_1$-vectors are parallel to the $c$-axis and hence, are unaffected themselves by any in-plane rotation of the magnetic field direction. Then, it is suggested that, based on the representation (3), the sudden switching of the ${\bm Q}$-vector upon rotating the in-plane field direction through (1,0,0) leads to the simulatneous change of the gap node of the induced spin-triplet vector ${\bm d}_1$. This is the same as the interpretation introduced \cite{Kim} to explain the thermal conductivity data. 

\begin{figure}[htbp]
 \begin{minipage}{0.5\hsize}
  \begin{center}
   \includegraphics[width=60mm]{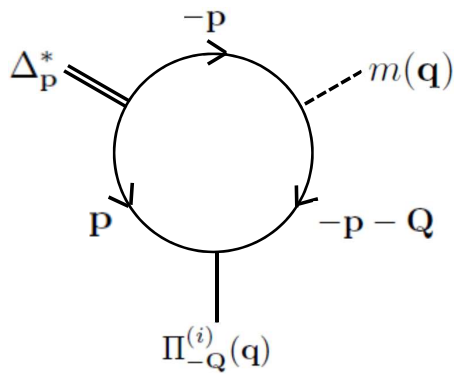}
\end{center}
 \end{minipage}
 \caption{Diagram describing the coupling term, eq. (4) or (5), among the three order parameters. Compare the wave vectors carried by each Green's functions (solid curves) with the indices indicated in eqs.(1)-(3). 
}
  \label{fig.2:}
\end{figure}

In deriving the third order free energy term (4) microscopically, however, 
the above representation on realization of a $\pi$-triplet order should be changed. In fact, the linearized representation such as $\psi({\bm p})$ and ${\bm d}_1({\bm p})$ is not useful for describing the ${\bm p}$-dependences of the pairing functions and the SDW order consistently, and they have to be rewritten in the tight-binding representation. The pairing functions need to be replaced, in the tight-binding model, in the manner 
\begin{eqnarray}
\psi({\bm p}) &\propto& p_x^2 - p_y^2 
\to w_{\bm p} = {\rm cos}(p_x) - {\rm cos}(p_y), \\ \nonumber 
{\bm d}_1({\bm p}) &\propto& {\hat {\bm z}} (p_x \mp p_y) \to {\tilde {\bm d}}_1({\bm p}) \propto {\hat {\bm z}} [{\rm sin}(p_x) \mp {\rm sin}(p_y)].
\end{eqnarray}
In addition, the diagram representation, Fig.2, of the coupling term (4) implies that, in order for this term to become nonzero, the $\pi$-triplet order parameter should be expressed by shifting ${\bm p}$ in eq.(3) to ${\bm p}+{\bm Q}/2$ in the form consistent with the expressions (1) and (2) of other order parameters. That is, if the alternative representation of the $\pi$-triplet order 
parameter 
\begin{equation}
{\it \Pi}^{(s)}_{\pm{\bm Q}} = \frac{V_s}{2} \sum_{\alpha, \beta, {\bm p}} {\bm D}_{s}({\bm p}) \cdot (-i \bm{\sigma} \sigma_y)_{\alpha, \beta} \langle c_{{\bm p},\alpha} c_{-{\bm p} \pm {\bm Q}, \beta} \rangle.
\end{equation}
is used to obtain the free energy, the pairing function in the $\Gamma_1$-representation is given by 
\begin{equation}
{\bm D}_{1}({\bm p}) = {\tilde {\bm d}}_{1}({\bm p}\mp{\bm Q}/2) \propto {\hat {\bm z}} w_{\bm p}
\end{equation}
{\it irrespective of} the ${\bm Q}$-direction, where $w_{\bm p}$ is defined in eq.(6). Thus, there is no change of gap nodes of the triplet order parameter accompanying the discontinuous change of the SDW ${\bm Q}$-direction in the tight-binding representation. Note that ${\it \Pi}^{(s)}$ and ${\tilde {\it \Pi}}^{(s)}$ are defined by summing over the momentum ${\bm p}$ so that they are equivalent to each other. Nevertheless, in examining the free energy and the resulting phase diagram, this tight-binding representation eq.(7) has to be used to make our calculation consistent with the conventional definition of other order parameters, eqs.(1) and (2). On the other hand, the Doppler shift to be examined in relation to the the thermal conductivity data \cite{Kim} 
is investigated based on the use of the continuum representation, eq.(3) 
(see also sec.VI). 

\begin{figure}[htbp]
 \begin{minipage}{0.5\hsize}
  \begin{center}
   \includegraphics[width=60mm]{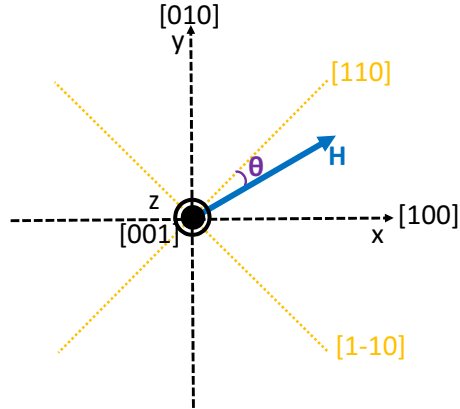}
\end{center}
 \end{minipage}
 \caption{Definition of coordinates in the $a$-$b$ (basal) plane. The direction of the magnetic field applied along the basal plane is expressed by the angle $\theta$ measured from the (1,1,0) direction. 
}
  \label{fig.3}
\end{figure}

The $\pi$-triplet order parameter making another coupling term (5) nonvanishing can similarly be considered by noting that the magnetic field is perpendicular to the SDW moment \cite{Kenzel} parallel to the $c$-axis, and one finds that any ${\bm d}_3$ and ${\bm d}_4$ belonging to $\Gamma_3$ or $\Gamma_4$ in Table I satisfy this condition. According to eqs.(6) to (8), the two order parameters ${\bm D}_3$ belonging to $\Gamma_3$ in the tight-binding approximation are given by $({\rm cos}p_x, \mp{\rm cos}p_y, 0)$ and $({\rm cos}p_y, \mp{\rm cos}p_x, 0)$ for ${\bm Q}=(k, \pm k, 0.5)$, respectively, while the corresponding ones ${\bm D}_4$ in $\Gamma_4$ are given by $({\rm cos}p_x, \pm{\rm cos}p_y, 0)$ and $({\rm cos}p_y, \pm{\rm cos}p_x, 0)$. All of them are gapped in the ${\bm Q}$-directions parallel to the $d$-wave gap nodes, in disagreement with the indication of the thermal conductivity data \cite{Kim}. The phase diagrams following from them will be discussed at the end of the next section. We note that both ${\bm D}_3$ and ${\bm D}_4$ are perpendicular to the $c$-axis and thus, in contrast to ${\bm D}_1$, depend on the in-plane direction of the magnetic field perpendicular to the $c$-axis. In fact, when the magnetic field field ${\bm H}$ is tilted within the $a$-$b$ plane up to the angle $\theta$ from (1,1,0), as defined in Fig.3, the parallel component $D_s(\theta)|_\parallel$ and perpendicular one $D_s(\theta)|_\perp$ to ${\bm H}$ of ${\bm D}_s = (D_{s,x}, D_{s,y},0)$ ($s=3$ and $4$) are given as the following $\theta$-dependent expressions; 
\begin{eqnarray}
D_s(\theta)|_\parallel &=& \frac{1}{\sqrt{2}} [ (D_{s,x} + D_{s,y}){\rm cos}\theta + (D_{s,x} - D_{s,y}){\rm sin}\theta ], \nonumber \\
D_s(\theta)|_\perp &=& \frac{1}{\sqrt{2}} [ (D_{s,y} - D_{s,x}){\rm cos}\theta + (D_{s,x} + D_{s,y}){\rm sin}\theta ], 
\end{eqnarray}
respectively. 

\section{Model and stable ${\it \Pi}$-triplet order}

The recent thermal conductivity data \cite{Kim} have suggested the presence of a $\pi$-triplet order with gap nodal lines perpendicular to the basal plane. The results in sec.I indicate that the realized $\pi$-triplet order should be not ${\bm D}_s$ ($s=3$ and $4$) but ${\bm D}_1$ defined in eq.(8). It will be shown here that, indeed, the $\bm{D}_1$ state tends to have a lower free energy. 

Throughout this paper, we focus on the Pauli-limited model with no orbital pair-breaking effect included. That is, the presence of the field-induced vortices will be neglected. This approximation which has also been used elsewhere \cite{other,Hatakefinal,Hosoya} seems to give quantitatively reasonable results as far as the mean field approximation is used to describe the phase diagram. 

First, we start from describing the model to be used in sec.V where all of the SDW, FFLO, and $\pi$-triplet orders are taken into account. The following mean field Hamiltonian is essentially the same as that broadly used in the literature \cite{Aperis} and expressed as 

\begin{eqnarray}
{\cal H}={\cal H}_{0}+{\cal H}_{\rm SS}+{\cal H}_{\rm SDW}+{\cal H}_{\rm TS},
\end{eqnarray} 
where ${\cal H}_{0}$ is the sum of the transfer energy and the Zeeman term, 
\begin{eqnarray}
{\cal H}_{\rm 0} &=& \sum^{}_{\sigma=\pm 1}\int_{}^{ }d^{3} {\bm r}   [\psi^{(\sigma)}]^\dagger({\bm r})
\Bigl[\varepsilon(- i {\bm \nabla})-\sigma I \Bigr] \psi^{(\sigma)}({\bm r}),\nonumber\\
\psi^{(\sigma)}({\bm r}) &=& \frac{1}{\sqrt{V}}\sum^{}_{\bm{p}} c_{{\bm p},\sigma}e^{i{\bm p} \cdot {\bm r}},\nonumber\\
\varepsilon({\bm p}) &=& -2t_{1}(\cos(p_{x})+\cos(p_{y}))-4t_{2}\cos(p_{x})\cos(p_{y}) \nonumber\\
&-& 2t_{3}(\cos(2p_{x})+\cos(2p_{y}))-\mu,
\end{eqnarray}
and, following the previous study \cite{Hosoya} neglecting the $\pi$-triplet order, the parameter values $t_{1}/T_{c}=15$, $t_{2}/t_{1}=-1.5$, $t_{3}/t_{1}=-0.65$, and $\mu/t_{1}=1.85$ have been used. Taking account of the case with a spatial modulation of the $d$-wave order parameter ${\it \Delta}$, the second term of eq.(10) associated with the $d$-wave SC pairing will be expressed in the form
\begin{eqnarray}
{\cal H}_{\rm SS}=
 \sum_{{\bm q}_{\rm LO}} \biggl[ \frac{1}{|g|} |{\it \Delta}({\bm q}_{\rm LO})|^{2} 
-\left( {\it \Delta}({\bm q}_{\rm LO}) \Psi^{\dag}({\bm q}_{\rm LO}) +{\rm H.c.} \right) \biggr],
\end{eqnarray} 
where 
\begin{eqnarray}
\Psi({\bm q}_{\rm LO}) &=& \frac{1}{2} \sum^{}_{{\bm p},\alpha,\beta} (-i \sigma_{y})_{\alpha,\beta} \, w_{\bm p} \, 
c_{-{\bm p}+{\bm q}_{\rm LO}/2,\alpha} \, c_{{\bm p}+{\bm q}_{\rm LO}/2,\beta}, \nonumber \\ 
{\it \Delta}({\bm q}_{\rm LO}) &=& |g|\langle \Psi({\bm q}_{\rm LO}) \rangle. 
\end{eqnarray}
That is, a possible FFLO spatial modulation with the wave vector ${\bm q}_{\rm LO}$ of the $d$-wave SC order parameter is included in the above expressions. 
Then, noting that, in the present issue, the SDW ordering occurs in the SC phase, the SDW order parameter should also be generally ${\bm q}_{\rm LO}$-dependent. Thus, the SDW mean field part of the Hamiltonian is expressed by the term 
\begin{eqnarray}
{\cal H}_{\rm SDW}= \sum^{}_{{\bm q},{\bm q}_{\rm LO}} \biggl[ U^{-1} |m({\bm q},{\bm q}_{\rm LO})|^{2}
- \left(m({\bm q},{\bm q}_{\rm LO}) S^{\dag}({\bm q},{\bm q}_{\rm LO}) +{\rm H.c.} \right) \biggr], 
\end{eqnarray} 
where ${\bm q}$ indicates possible incommensurate components ${\bm Q} - {\bm Q}_0$, and 
\begin{eqnarray}
S({\bm q},{\bm q}_{\rm LO}) &=& \sum^{}_{{\bm p},\alpha,\beta} 
c_{{\bm p},\alpha}^{\dag} \,
(\bm{\sigma} \cdot \hat{\bm{n}})_{\alpha,\beta} \, 
c_{{\bm p}+{\bm Q}+{\bm q}_{\rm LO},\beta}, \nonumber \\ 
m({\bm q},{\bm q}_{\rm LO}) &=& U\langle S({\bm q},{\bm q}_{\rm LO}) \rangle .
\end{eqnarray}
Hereafter, the $z$-axis will be chosen along the magnetic field ${\bm H}$ in the {\it spin} space. Then, to study the HFLT phase of CeCoIn$_5$ with the SDW moment along the $c$-axis and hence, perpendicular to ${\bm H}$, $\bm{n}$ in eq.(15) will be taken to be along the $y$-axis. 

Further, we assume the presence of a weakly attractive channel for the $\pi$-triplet pairing with the interaction strength $V_1$ ($ > 0$). The terms associated with the triplet pairing component expressed by ${\cal H}_{\rm TS}$ take the form 
\begin{eqnarray}
{\cal H}_{\rm TS}
&=& \frac{1}{V_{1}}\sum^{}_{{\bm q},{\bm q}_{\rm LO}}
\left(|{\it \Pi}^{(1)}_{-{\bm Q}}({\bm q},{\bm q}_{\rm LO})|^{2} + |{\it \Pi}^{(1)}_{{\bm Q}}({\bm q},{\bm q}_{\rm LO})|^{2} \right)
\nonumber \\ 
&-&\sum^{}_{{\bm q},{\bm q}_{\rm LO}}\left({\it \Pi}^{(1)}_{-{\bm Q}}({\bm q},{\bm q}_{\rm LO}) 
B^{(1)  \dag}_{-{\bm Q}}({\bm q},{\bm q}_{\rm LO}) 
+ {\it \Pi}^{(1)}_{\bm Q}({\bm q},{\bm q}_{\rm LO}) B^{(1) \dag}_{\bm Q}({\bm q},{\bm q}_{\rm LO}) 
+{\rm H.c.} \right)
\end{eqnarray} 
in the case of the $\pi$-triplet order ${\bm D}_1$, where 
\begin{eqnarray}
B^{(1)}_{-{\bm Q}}({\bm q},{\bm q}_{\rm LO}) &=& \frac{1}{2} \sum^{}_{{\bm p},\alpha,\beta} 
(-i {\bm D}_{1}({\bm p}) \cdot
{\bm{\sigma}}, \sigma_{y})_{\alpha,\beta} \, 
c_{{\bm p},\alpha} \,
c_{-{\bm p}-{\bm Q}- {\bm q}_{\rm LO},\beta}, \nonumber \\ 
B^{(1)}_{\bm Q}({\bm q},{\bm q}_{\rm LO}) &=& \frac{1}{2} \sum^{}_{{\bm p},\alpha,\beta} 
(-i {\bm D}_{1}({\bm p}) \cdot
{\bm{\sigma}}, \sigma_{y})_{\alpha,\beta} \, 
c_{{\bm p}+{\bm Q}+{\bm q}_{\rm LO},\alpha} \, 
c_{-{\bm p},\beta}, \nonumber \\ 
{\it \Pi}^{(1)}_{-{\bm Q}}({\bm q},{\bm q}_{\rm LO}) &=& V_1 \langle B^{(1)}_{-{\bm Q}}
({\bm q},{\bm q}_{\rm LO}) \rangle , \nonumber \\ 
{\it \Pi}^{(1)}_{\bm Q}({\bm q},{\bm q}_{\rm LO}) &=& V_1 \langle B^{(1)}_{\bm Q}
({\bm q},{\bm q}_{\rm LO}) \rangle. 
\end{eqnarray} 

In the present and next sections, we will not treat the full Hamiltonian ${\cal H}$. To understand which of the triplet-pairings induced by the SDW order is stable, the possible FFLO spatial modulation of the $d$-wave SC order parameter ${\it \Delta}$ will be neglected for a while so that we focus on the ${\bm q}_{\rm LO}=0$ term in ${\cal H}$. Hereafter, ${\it \Pi}^{(n)}_{\bm Q}({\bm q},{\bm q}_{\rm LO}=0)$ and $m({\bm q},{\bm q}_{\rm LO}=0)$ will simply be written as ${\it \Pi}^{(n)}_{\bm Q}({\bm q})$ and $m({\bm q})$, respectively. Then, the free energy density following from our calculation in this section is divided into three terms 
\begin{eqnarray}
f = f_{{\it \Delta}, (0)}+f_{m}+f_{{\it \Pi}}. 
\label{eq:free_pi1}
\end{eqnarray}
In our Pauli-limited treatment, the $d$-wave SC order parameter ${\it \Delta}$ can be included fully in $f$ through the formula \cite{Gert} 
\begin{eqnarray}
f_{\it \Delta}(q_{\rm LO})
= 
 \frac{|{\it \Delta}|^{2}}{|g|} 
+ \frac{T}{2}
\sum^{\infty}_{\varepsilon_{n}=-\infty} \sum^{}_{{\bm p},\sigma} \int_{\varepsilon_{n}}^{\infty s_{\epsilon}}
\!d\omega\  \mathop{\mathrm{Tr}}
\biggl[ i \sigma_{z} {\hat G}^{(\sigma)}_{\omega,(0)}({\bm p}) \biggr].
\label{eq:fdl}, 
\end{eqnarray}
where $s_\epsilon=\varepsilon/|\varepsilon|$, and ${\hat G}^{(\sigma)}_{\omega,(0)}$ denotes the Gor'kov Green's function defined in the manner 
\begin{eqnarray}
{\hat G}^{(\sigma)}_{\varepsilon_{n}, (0)}({\bm p}) &=&
\left[
\begin{array}{cc}
i\varepsilon_{n} - \varepsilon({\bm p}) + \sigma I & - \sigma {\it \Delta}_{\bm p} \\
\sigma {\it \Delta}^{\ast}_{\bm p} & -i\varepsilon_{n} - \varepsilon({\bm p}) - \sigma I \\
\end{array} 
\right]^{-1} \nonumber\\
&=&
\frac{1}{\varepsilon({\bm p})^{2}-(i\varepsilon_{n}+\sigma I)^{2}+|{\it \Delta}_{\bm p}|^{2}} \left[ 
\begin{array}{cc}
-i\varepsilon_{n} - \varepsilon({\bm p}) - \sigma I & \sigma {\it \Delta}_{\bm p} \\
- \sigma {\it \Delta}^{\ast}_{\bm p} & i\varepsilon_{n} - \varepsilon({\bm p}) + \sigma I \\
\end{array} 
\right] \nonumber\\
&=&
\left[ 
\begin{array}{cc}
G^{(\sigma)}_{\varepsilon_{n}}({\bm p}) & F^{(\sigma)}_{\varepsilon_{n}}({\bm p}) \\
{\overline F}^{(\sigma)}_{\varepsilon_{n}}({\bm p}) & {\overline G}^{(-\sigma)}_{\varepsilon_{n}}({\bm p}) \\
\end{array} 
\right] 
\end{eqnarray}
with $\Delta_{\bm p}=\Delta w_{\bm p}$. Using these expressions, the first term $f_{{\it \Delta}, (0)}$ of eq.(18) consisting only of ${\it \Delta}$ is easily rewritten as 
\begin{eqnarray}
f_{{\it \Delta},(0)} &=& \frac{|{\it \Delta}|^2}{|g|} - T \sum_{\varepsilon_n>0, {\bm p}}
	\ln \biggl[\frac{(\varepsilon_n^2+[\varepsilon({\bm p})]^2+|{\it \Delta}_{\bm p}|^2-I^2)^2+4 \varepsilon_n^2 I^2}{(\varepsilon_n^2+[\varepsilon({\bm p})]^2-I^2)^2+4 \varepsilon_n^2 I^2} \biggr]. \nonumber \\ 
\label{SC_part}
\end{eqnarray}

On the other hand, the second term $f_{m}$ of eq.(18) expresses the GL expansion in the SDW order parameter $m$, while $f_{{\it \Pi}}$ denotes additional terms occurring by taking account of the $\pi-$triplet SC order. They will be divided below into several terms like
\begin{eqnarray}
f_{m} &=& f^{(2)}_{m}+f^{(4)}_{m}, \nonumber\\
f_{{\it \Pi}} &=& f^{(1, 1)}_{{\it \Pi}, m}+f^{(2)}_{{\it \Pi}}. 
\label{eq:free_m_pi}
\end{eqnarray}
The coupling term (4) or (5) given in sec.I corresponds to the first term 
$f^{(1,1)}_{{\it \Pi},m}$ of $f_{\it \Pi}$. 
First, using the Green's functions, $f^{(2)}_{m}$ and $f^{(4)}_{m}$ take the form 
\begin{eqnarray}
f^{(2)}_{m} &=& \sum_{\bm q} 
\biggl[
\frac{1}{U} 
+ \frac{T}{2}
\sum^{\infty}_{\varepsilon_{n}=-\infty} \sum^{}_{{\bm p},\sigma, s_{1}=\pm 1}
{\rm Tr}
\biggl(
{\hat G}^{(\sigma)}_{\varepsilon_n, \, (0)}({\bm p}+{\bm Q}_{0}+s_{1}{\bm q})
{\hat G}^{(-\sigma)}_{\varepsilon_n, \, (0)}({\bm p})
\biggr)
\biggr]
|m({\bm q})|^{2} \\ \nonumber
&=& \sum_{\bm q} 
\biggl[
\frac{1}{U} 
- 4T
\sum^{}_{\varepsilon_{n}>0, {\bm p}} 
\frac{\varepsilon_{n}^{2}+I^{2} - \varepsilon({\bm p})\varepsilon({\bm p}+{\bm Q}) 
- {\it \Delta}^{\ast}_{\bm p} {\it \Delta}_{{\bm p}+{\bm Q}} }
{c_{1}^{2}+d_{1}^{2}} c_{1}
\biggr]
|m({\bm q})|^{2},
\nonumber \\  
\label{eq:free_m2}, 
\end{eqnarray}
and 
\begin{eqnarray}
f^{(4)}_{m} &=& 
\frac{T}{4}
\sum^{\infty}_{\varepsilon_{n}=-\infty} \sum^{}_{{\bm p},{\bm q}, \sigma, s_{1}=\pm 1} 
{\rm Tr}
\biggl(
{\hat G}^{(\sigma)}_{\varepsilon_n, \, (0)}({\bm p}+{\bm Q}_{0}+s_{1}{\bm q}) 
{\hat G}^{(-\sigma)}_{\varepsilon_n, \, (0)}({\bm p})\nonumber \\ 
&\times& \ 
{\hat G}^{(\sigma)}_{\varepsilon_n, \, (0)}({\bm p}+{\bm Q}_{0}+s_{1}{\bm q})
{\hat G}^{(-\sigma)}_{\varepsilon_n, \, (0)}({\bm p})
\biggr)
|m({\bm q})|^{4} \\ \nonumber
&=& 
2T
\sum^{}_{\varepsilon_{n}>0, {\bm p}, {\bm q}}
\frac{1}{ ( c_{1}^{2}+d_{1}^{2} )^{2} }
\bigl[ ( c_{1}^{2}-d_{1}^{2} ) [ ( \varepsilon_{n}^{2}+I^{2}-\varepsilon({\bm p})\varepsilon({\bm p}+{\bm Q}) 
- {\it \Delta}^{\ast}_{\bm p} {\it \Delta}_{{\bm p}+{\bm Q}} )^{2}  
\nonumber \\ 
&-& \varepsilon_{n}^{2} (  ( \varepsilon({\bm p}) + \varepsilon({\bm p}+{\bm Q}) )^{2} 
+ |{\it \Delta}_{\bm p} + {\it \Delta}_{{\bm p}+{\bm Q}} |^{2}    )
+ I^{2} (  ( \varepsilon({\bm p}) - \varepsilon({\bm p}+{\bm Q}) )^{2}  \nonumber \\ 
&+& | {\it \Delta}_{\bm p} - {\it \Delta}_{{\bm p}+{\bm Q}} |^{2}) 
- | {\it \Delta}_{\bm p} \varepsilon({\bm p}+{\bm Q}) 
- {\it \Delta}_{{\bm p}+{\bm Q}} \varepsilon({\bm p}) 
|^{2}
]
- 2 c_{1} d_{1}^{2} 
\bigr]
|m({\bm q})|^{4}.
\nonumber \\ 
\label{eq:free_m4}
\end{eqnarray}
The coefficients $c_1$ and $d_1$ will be defined later. 

Next, we turn to $f_{\it \Pi}$. In the case with the triplet pairing belonging to the irreducible representation $\Gamma_1$, the first term $f^{(1, 1)}_{{\it \Pi}, m}$ 
of the free energy $f_{\it \Pi}$ associated with the $\pi$-triplet pairing is expressed in terms of ${\bm D}_1 = {\hat {\bm z}} w_{\bm p}$ as 
\begin{eqnarray}
f^{(1, 1)}_{{\it \Pi}, m} &=& 
\frac{T}{2}
\sum^{\infty}_{\varepsilon_{n}=-\infty} 
\sum^{}_{{\bm p}, {\bm q}, \sigma, s_{1}=\pm 1} 
{\rm Tr}
\biggl( \sum^{}_{j} 
\sigma A_{j}^{(1)} \hat{a}_{j}^{(1)}
{\hat G}^{(\sigma)}_{\varepsilon_n, \, (0)}({\bm p}+{\bm Q}_{0}+s_{1}{\bm q})
\hat{b}_{j}^{(1)}
{\hat G}^{(-\sigma)}_{\varepsilon_n, \, (0)}({\bm p})
B_{j}^{(1)}
+ {\rm H.c.}
\biggr),
\nonumber \\ 
\label{eq:Green_d1_11}
\end{eqnarray}
where $\sigma_{0}=1$, $\hat{e}_{12} = (\sigma_x + i \sigma_y)/2$, and the coefficients $A_{j}^{(1)}$, $\hat{a}_{j}^{(1)}$ , $\hat{b}_{j}^{(1)}$, and $B_{j}^{(1)}$ are defined in Table \ref{tb:d1_11} below 

\begin{table}[htb]
\begin{center}
\caption{Definition of the coefficients $A_{j}^{(1)}$, $\hat{a}_{j}^{(1)}$ , $\hat{b}_{j}^{(1)}$, and $B_{j}^{(1)}$ in eq.(\ref{eq:Green_d1_11}).}
  \begin{tabular}{ccccc} \hline
    $j$  & $A_{j}^{(1)}$ & $\hat{a}_{j}^{(1)}$ & $\hat{b}_{j}^{(1)}$ & $B_{j}^{(1)}$ \\ \hline
    $1$  & $-w_{\bm p}$ & $\hat{e}_{12}$ & $\sigma_{0}$ 
    & $m({\bm q}) {\it \Pi}^{(1)}_{-{\bm Q}}({\bm q}) $ \\
    $2$  & $w_{\bm p}$ & $\sigma_{0}$ & $\hat{e}_{12}$ 
    & $m^{\ast}({\bm q}) {\it \Pi}^{(1)}_{\bm Q}({\bm q}) $ \\ \hline
  \end{tabular}
  \label{tb:d1_11}
  \end{center}
\end{table}

The second term $f^{(2)}_{{\it \Pi}}$ of $f_{\it \Pi}$ is 
\begin{eqnarray}
f^{(2)}_{{\it \Pi}} &=& \sum_{\bm q} 
\frac{ |{\it \Pi}^{(1)}_{-{\bm Q}}({\bm q})|^{2} + |{\it \Pi}^{(1)}_{\bm Q}({\bm q})|^{2} }{V_1} \nonumber \\ 
&-& 
\frac{T}{2}
\sum^{\infty}_{\varepsilon_{n}=-\infty} 
\sum^{}_{{\bm p}, {\bm q}, \sigma, s_{1}=\pm 1} 
{\rm Tr}
\biggl( \sum_j 
{A'}_{j}^{(1)} \hat{a'}_{j}^{(1)}
{\hat G}^{(\sigma)}_{\varepsilon_n, \, (0)}({\bm p}+{\bm Q}_{0}+s_{1}{\bm q})
\hat{b'}_{j}^{(1)}
{\hat G}^{({\alpha'}_{j}^{(1)})}_{\varepsilon_n, \, (0)}({\bm p})
{B'}_{j}^{(1)} 
\biggr) ,
\nonumber \\ 
\label{eq:Green_d1_2}
\end{eqnarray}
where $\hat{e}_{21} = (\sigma_x - i \sigma_y)/2$, and the coefficients ${A'}_{j}^{(1)}$, $\hat{a'}_{j}^{(1)}$ , $\hat{b'}_{j}^{(1)}$, ${\alpha'}_{j}^{(1)}$, and ${B'}_{j}^{(1)}$ are given in Table \ref{tb:d1_2}

\begin{table}[htb]
\begin{center}
\caption{Definition of the coefficients ${A'}_{j}^{(1)}$, $\hat{a'}_{j}^{(1)}$, $\hat{b'}_{j}^{(1)}$, ${\alpha'}_{j}^{(1)}$, and ${B'}_{j}^{(1)}$ in eq.(\ref{eq:Green_d1_2}).}
  \begin{tabular}{cccccc} \hline
    $j$  & ${A'}_{j}^{(1)}$ & $\hat{a'}_{j}^{(1)}$ & $\hat{b'}_{j}^{(1)}$ & ${\alpha'}_{j}^{(1)}$ & ${B'}_{j}^{(1)}$ \\ \hline
    $1$  & $w_{\bm p}^{2}$ & $\hat{e}_{12}$ & $\hat{e}_{21}$ & $-\sigma$ 
    & $|{\it \Pi}^{(1)}_{-{\bm Q}}({\bm q})|^{2} $ \\
    $2$  & $w_{\bm p}^{2}$ & $\hat{e}_{21}$ & $\hat{e}_{12}$ & $-\sigma$ 
    & $|{\it \Pi}^{(1)}_{\bm Q}({\bm q})|^{2} $ \\
    $3$  & $w_{\bm p}^{2}$ & $\hat{e}_{12}$ & $\hat{e}_{12}$ & $-\sigma$
    & ${\it \Pi}^{(1)}_{-\bm Q}({\bm q}) {\it \Pi}^{(1)}_{\bm Q}({\bm q})$ \\
    $4$  & $w_{\bm p}^{2}$ & $\hat{e}_{21}$ & $\hat{e}_{21}$ & $-\sigma$
    & ${\it \Pi}^{(1)\ast}_{-{\bm Q}}({\bm q}) {\it \Pi}^{(1)\ast}_{\bm Q}({\bm q})$ \\ \hline
  \end{tabular}
  \label{tb:d1_2}
  \end{center}
\end{table}

Rewriting eqs.(\ref{eq:Green_d1_11}) and (\ref{eq:Green_d1_2}), we have 
\begin{eqnarray}
f^{(1, 1)}_{{\it \Pi}, m}
&=&  
2T \sum^{}_{\varepsilon_{n}>0, {\bm p}, {\bm q}}
\frac{c_{1}}{ c_{1}^{2}+d_{1}^{2} }
\biggl((\varepsilon({\bm p}) {\it \Delta}^{\ast}_{{\bm p}+{\bm Q}} 
- \varepsilon({\bm p}+{\bm Q}){\it \Delta}^{\ast}_{\bm p})  \nonumber \\ 
&\times& (- m({\bm q}) w_{\bm p} {\it \Pi}^{(1)}_{-{\bm Q}}({\bm q}) 
+ m^{\ast}({\bm q}) w_{\bm p} {\it \Pi}^{(1)}_{\bm Q}({\bm q}) )
+ {\rm H.c.} \biggr) ,
\label{eq:d1_free11}
\end{eqnarray}
\begin{eqnarray}
f^{(2)}_{{\it \Pi}}
&=& \sum_{\bm q} 
\frac{ |{\it \Pi}^{(1)}_{-{\bm Q}}({\bm q})|^{2} + |{\it \Pi}^{(1)}_{\bm Q}({\bm q})|^{2} }{V_1} \nonumber \\ 
&-& 2T \sum^{}_{\varepsilon_{n}>0, {\bm p}, {\bm q}}
\frac{c_{1}}{ c_{1}^{2}+d_{1}^{2} } 
\biggl( \varepsilon_{n}^{2}+I^{2} + \varepsilon({\bm p})\varepsilon({\bm p}+{\bm Q}) \biggr)
\biggl( |w_{\bm p} {\it \Pi}^{(1)}_{-{\bm Q}}({\bm q})|^{2} 
+ |w_{\bm p} {\it \Pi}^{(1)}_{\bm Q}({\bm q})|^{2} \biggr) \nonumber \\ 
&+& 2T \sum^{}_{\varepsilon_{n}>0, {\bm p}, {\bm q}}
w_{\bm p}^{2}
\biggl( \frac{c_{1}}{ c_{1}^{2}+d_{1}^{2} }
{\it \Delta}^{\ast}_{\bm p} {\it \Delta}^{\ast}_{{\bm p}+{\bm Q}}
w_{\bm p} {\it \Pi}^{(1)}_{-{\bm Q}}({\bm q}) 
w_{\bm p} {\it \Pi}^{(1)}_{\bm Q}({\bm q}) 
+ {\rm H.c.} \biggr).
\label{eq:d1_free2}
\end{eqnarray}
Here, 
\begin{eqnarray}
a_1 &=& [\varepsilon({\bm p})]^2+\varepsilon_n^2+|{\it \Delta}_{\bm p}|^2-I^2, \nonumber \\
a_2 &=& [\varepsilon({\bm p}+{\bm Q})]^2
+\varepsilon_n^2+|{\it \Delta}_{{\bm p}+{\bm Q}}|^2-I^2, \nonumber \\
b_1 &=& 2 \varepsilon_n I, \nonumber \\
c_{1} &=& a_{1}a_{2} + b_{1}^{2}, \nonumber \\ 
d_{1} &=& b_{1} ( a_{2} - a_{1} ) .
\label{eq:abcd}
\end{eqnarray}

Effects of the $\pi$-triplet order on the free energy can be incorporated by minimizing $f_{\it \Pi}$ with respect to the $\pi$-triplet order parameters ${\it \Pi}^{(1)}_{\pm {\bm Q}}$. The resulting ${\it \Pi}^{(1)}_{\pm {\bm Q}}$ is proportional to the SDW order parameter and given by 
\begin{eqnarray}
{\it \Pi}^{(1)}_{-{\bm Q}}({\bm q}) &=& 
\frac{-2T \sum_{\varepsilon_n > 0, {\bm p}} 
(\varepsilon({\bm p}) {\it \Delta}^{\ast}_{{\bm p}+{\bm Q}} 
- \varepsilon({\bm p}+{\bm Q}) {\it \Delta}^{\ast}_{\bm p} ) c_1 
w_{\bm p}/(c_1^2 + d_1^2)}
{ V_1^{-1} 
- 2T \sum_{\varepsilon > 0, {\bm p}} 
( \varepsilon_{n}^{2}+I^{2} + \varepsilon({\bm p})\varepsilon({\bm p}+{\bm Q}) 
+ {\it \Delta}_{\bm p} {\it \Delta}^{\ast}_{{\bm p}+{\bm Q}} )
w_{\bm p}^{2} c_{1}/( c_{1}^{2}+d_{1}^{2} ) } 
m({\bm q}) \nonumber \\
&=& - {\it \Pi}^{(1)}_{\bm Q}({\bm q}).
\nonumber \\
\label{eq:pi_min}
\end{eqnarray}
By substituing this into $f_{\it \Pi}$, additional terms proportional to $|m|^2$ are created which change, e.g., the field range of the HFLT phase. 

In Fig.4, an example of the resulting phase diagram is shown. Within the parameter values used in our numerical computations, the coefficient of the $|m|^4$ term, eq.(26), is always positive. Thus, a second order transition signaling the appearance of a nonzero $|m|$ occurs on the lower (red) solid curve $H^*(T)$. Namely, $|m|^2$ is proportional to $H-H^*$ just above $H^*(T)$. 
Further, within the parameter values we have chosen, the denominator of eq.(\ref{eq:pi_min}) remains positive so that no nonvanishing $\pi$-triplet order occurs without the presence of the SDW order. Nevertheless, the $\pi$-triplet order induced by the SDW order is found to broaden the HFLT phase. 

\begin{figure}[htbp]
 \begin{minipage}{0.5\hsize}
  \begin{center}
   \includegraphics[width=60mm]{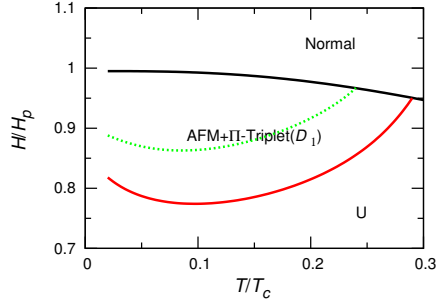}
\end{center}
 \end{minipage}
 \caption{HFLT phase in the field ($H$)-temperature ($T$) phase diagram in the Pauli-limited case with {\it no} FFLO order assumed. The used values of the coupling constants are $T_{c}/U=0.01622$ and $T_{c}/V_1=0.0075$, where $T_c$ is the SC transition temperature in zero field, and $H_p$ is the Pauli-limiting field at $T=0$. The upper (black) and the lower (red) solid curves are the first order mean field SC transition curve $H_{c2}(T)$ and the field-induced second order transition line $H^*(T)$ above which the HFLT phase with the $\Gamma_1$-triplet SC order is present, respectively, and the green dotted curve is the $H^*(T)$ line in the $V_1 \to 0$ limit. 
}
  \label{fig.4}
\end{figure}

Although the $\Gamma_1$-representation is a candidate of the staggered triplet order in the HFLT phase, this ${\bm d}_s$-vector is parallel to the $c$-axis and thus, cannot change under an in-plane rotation of the magnetic field ${\bm H}$ perpendicular to the $c$-axis. That is, an element neglected in this section needs to be taken into account to explain the switching, detected \cite{Gerber} in the neutron scattering measurement, of the SDW ${\bm Q}$-vector sensitive to ${\bm H}$. In the next section, we show that the FFLO order neglected in this section leads to the switching of the SDW ${\bm Q}$-vector upon the in-plane rotation of the magnetic field direction. 

Before ending this section, the resulting phase diagrams in the case with the $\pi$-triplet order ${\bm D}_s$ with $s=3$ or $4$ will be discussed. In this case, the original expression of the coupling term corresponding to eq.(5) is complicated and takes the form 

\begin{eqnarray}
f^{(1, 1)}_{{\it \Pi}, m} &=& 
- \frac{i \, T}{2\sqrt[]{\mathstrut 2}}
\sum^{\infty}_{\varepsilon_{n}=-\infty} \sum^{}_{{\bm p}, {\bm q}, \sigma, s_{1}=\pm 1} 
{\rm Tr}
\biggl( \sum^{}_{j} 
A_{j}^{(s)} \hat{a}_{j}^{(s)}
{\hat G}^{(\sigma)}_{\varepsilon_n, \, (0)}({\bm p}+{\bm Q}_{0}+s_{1}{\bm q})
\hat{b}_{j}^{(s)}
{\hat G}^{(-\sigma)}_{\varepsilon_n, \, (0)}({\bm p})
B_{j}^{(s)}
- {\rm H.c.}
\biggr) ,
\nonumber \\
\label{eq:Green_d3_11}
\end{eqnarray}
where the expressions of $A_{j}^{(s)}$, $\hat{a}_{j}^{(s)}$ , $\hat{b}_{j}^{(s)}$, and $B_{j}^{(s)}$ ($s=3$ or $4$) are defined in Table\ref{tb:d3_11}. 

\begin{table}[htb]
\begin{center}
\caption{Definition of the quantities $A_{j}^{(s)}$, $\hat{a}_{j}^{(s)}$ , $\hat{b}_{j}^{(s)}$, and $B_{j}^{(s)}$ in eq.(\ref{eq:Green_d3_11}). }
  \begin{tabular}{ccccc} \hline
    $j$  & $A_{j}^{(s)}$ & $\hat{a}_{j}^{(s)}$ & $\hat{b}_{j}^{(s)}$ & $B_{j}^{(s)}$ \\ \hline
    $1$  & $D_{s}(\theta)|_\perp$ & $\hat{e}_{12}$ & $\sigma_{0}$ 
    & $m({\bm q}) {\it \Pi}^{(s)}_{-{\bm Q}}({\bm q}) $ \\
    $2$  & $D_{s}(\theta)|_\perp$ & $\sigma_{0}$ & $\hat{e}_{12}$ 
    & $m^{\ast}({\bm q}) {\it \Pi}^{(s)}_{\bm Q}({\bm q}) $ \\ \hline
  \end{tabular}
  \label{tb:d3_11}
  \end{center}
\end{table}

Similarly, $f^{(2)}_{\it \Pi}$ in eq.(\ref{eq:free}) is expressed by 
\begin{eqnarray}
f^{(2)}_{\it \Pi} &=& \sum_{\bm q} 
\frac{ |{\it \Pi}^{(s)}_{-{\bm Q}}({\bm q})|^{2} + |{\it \Pi}^{(s)}_{\bm Q}({\bm q})|^{2} }{V_s} \nonumber \\ 
&-& 
\frac{T}{4}
\sum^{\infty}_{\varepsilon_{n}=-\infty} \sum^{}_{{\bm p}, {\bm q}, \sigma, s_{1}=\pm 1} 
{\rm Tr}
\biggl( \sum_j 
{A'}_{j}^{(s)} \hat{a'}_{j}^{(s)}
{\hat G}^{(\sigma)}_{\varepsilon_n, \, (0)}({\bm p}+{\bm Q}_{0}+s_{1}{\bm q})
\hat{b'}_{j}^{(s)}
{\hat G}^{({\alpha'}_{j}^{(s)} )}_{\varepsilon_n, \, (0)}({\bm p})
{B'}_{j}^{(s)} 
\biggr),
\label{eq:Green_d3_2}
\end{eqnarray}
where ${A'}_{j}^{(s)}$, $\hat{a'}_{j}^{(s)}$ , $\hat{b'}_{j}^{(s)}$, ${\alpha'}_{j}^{(s)}$, and ${B'}_{j}^{(s)}$ are defined in Table\ref{tb:d3_2}. 

\begin{table}[htb]
\begin{center}
\caption{Definition of the quantities ${A'}_{j}^{(s)}$, $\hat{a'}_{j}^{(s)}$ , $\hat{b'}_{j}^{(s)}$, ${\alpha'}_{j}^{(s)}$, and ${B'}_{j}^{(s)}$ in eq.(\ref{eq:Green_d3_2}).}
  \begin{tabular}{cccccc} \hline
    $j$  & $A_{j}^{(1)}$ & $\hat{a}_{j}^{(1)}$ & $\hat{b}_{j}^{(1)}$ & ${\alpha'}_{j}^{(s)}$ & $B_{j}^{(1)}$ \\ \hline
    $1$  & $[D_{s}(\theta)|_\perp]^2$ & $\hat{e}_{12}$ & $\hat{e}_{21}$ & $-\sigma$  & $|{\it \Pi}^{(s)}_{-{\bm Q}}({\bm q})|^{2} $ \\
    $2$  & $[D_{s}(\theta)|_\parallel]^2$ & $\hat{e}_{12}$ & $\hat{e}_{21}$ & $\sigma$ 
    & $|{\it \Pi}^{(s)}_{-{\bm Q}}({\bm q})|^{2} $ \\
    $3$  & $[D_{s}(\theta)|_\perp]^2$ & $\hat{e}_{21}$ & $\hat{e}_{12}$ & $-\sigma$
    & $|{\it \Pi}^{(s)}_{\bm Q}({\bm q})|^{2} $ \\
    $4$  & $[D_{s}(\theta)|_\parallel]^2$ & $\hat{e}_{21}$ & $\hat{e}_{21}$ & $\sigma$
    & $|{\it \Pi}^{(s)}_{\bm Q}({\bm q})|^{2} $ \\ 
    $5$  & $[D_{s}(\theta)|_\perp]^2$ & $\hat{e}_{12}$ & $\hat{e}_{12}$ & $-\sigma$ 
    & ${\it \Pi}^{(s)}_{-{\bm Q}}({\bm q}) {\it \Pi}^{(s)}_{\bm Q}({\bm q})$ \\
    $6$  & $-[D_{s}(\theta)|_\parallel]^2$ & $\hat{e}_{12}$ & $\hat{e}_{12}$ & $\sigma$ 
    & ${\it \Pi}^{(s)}_{-{\bm Q}}({\bm q}) {\it \Pi}^{(s)}_{\bm Q}({\bm q})$ \\
    $7$  & $[D_{s}(\theta)|_\perp]^2$ & $\hat{e}_{21}$ & $\hat{e}_{21}$ & $-\sigma$
    & ${\it \Pi}^{(s)\ast}_{-{\bm Q}}({\bm q}) {\it \Pi}^{(s)\ast}_{\bm Q}({\bm q})$ \\
    $8$  & $-[D_{s}(\theta)|_\parallel]^2$ & $\hat{e}_{21}$ & $\hat{e}_{21}$ & $\sigma$
    & ${\it \Pi}^{(s)\ast}_{-{\bm Q}}({\bm q}) {\it \Pi}^{(s)\ast}_{\bm Q}({\bm q})$ \\ \hline
  \end{tabular}
  \label{tb:d3_2}
  \end{center}
\end{table}
\begin{figure}[htbp]
 \begin{minipage}{0.45\hsize}
  \begin{center}
   \includegraphics[width=60mm]{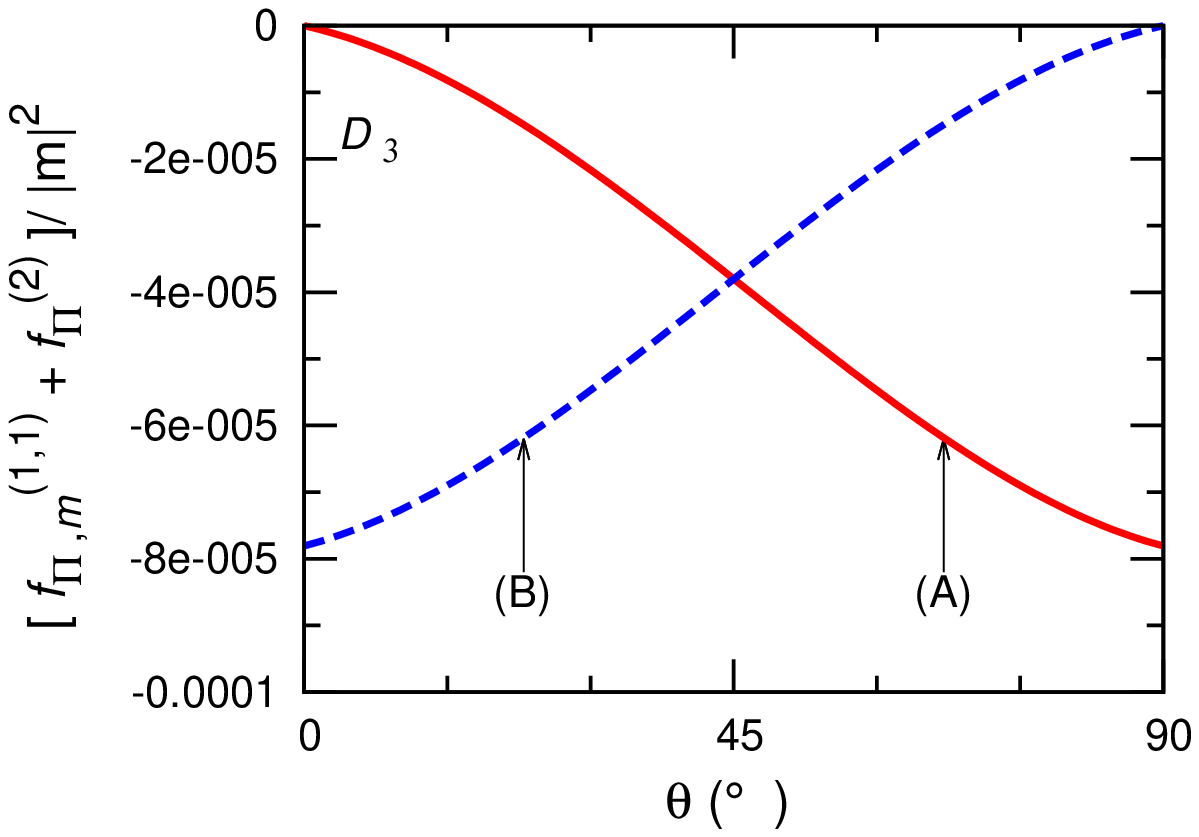}
  \hspace{1.6cm} (a)
\end{center}
 \end{minipage}
 \begin{minipage}{0.45\hsize}
  \begin{center}
   \includegraphics[width=60mm]{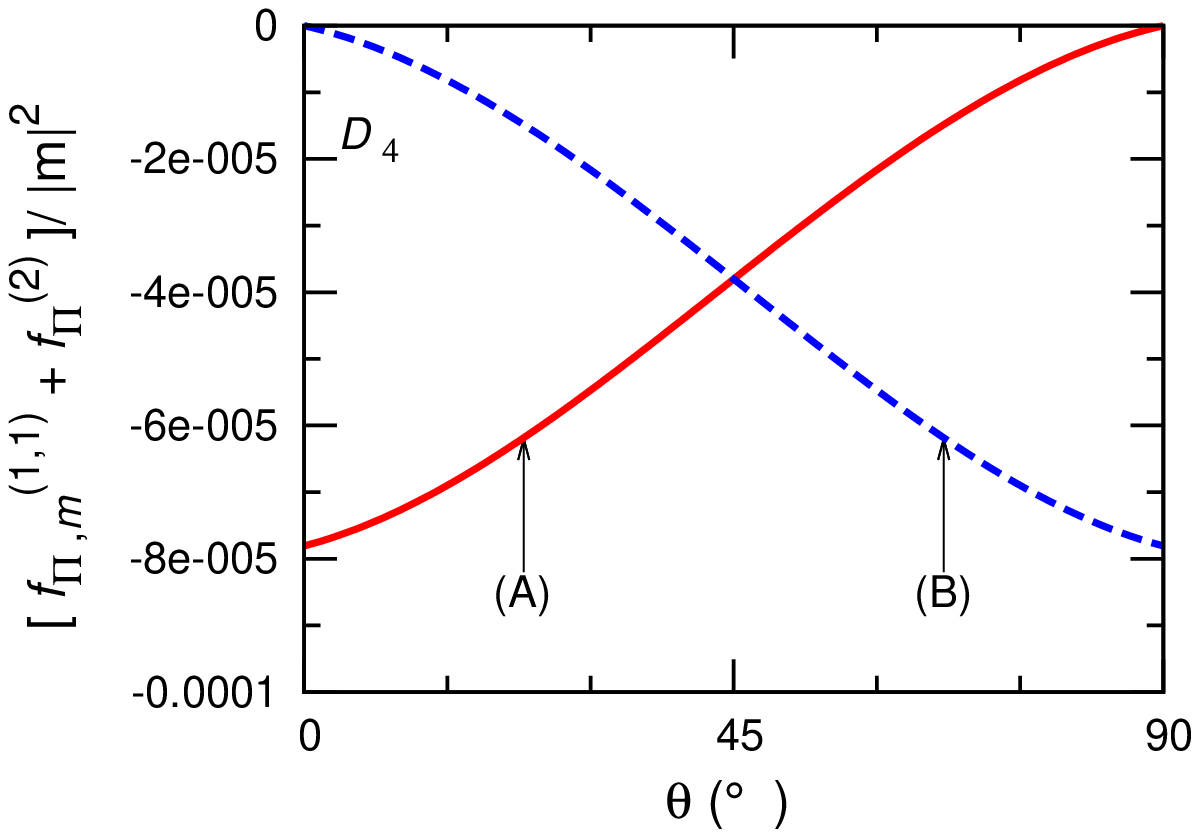}
  \hspace{1.6cm} (b)
\end{center}
 \end{minipage}
 \caption{ Normalized free energy $f^{(1, 1)}_{{\it \Pi}, m}+f^{(2)}_{\it \Pi}$ in the $\Gamma_3$ representation (a) and $\Gamma_4$ representation (b) obtained by using the parameters $H=0.91 H_{c}$, $T=0.1 T_{c}$, and $T_{c}/V_{s}=0.0033$. The solid (dashed) curve is the result in ${\bm Q} \parallel (1,1,0)$ (${\bm Q} \parallel (1,-1,0)$) case. For instance, when $0 \leq \theta \leq \pi/4$, the free energy with ${\bm Q} \parallel (1,-1,0)$ is lowered in the $\Gamma_3$ representation, while that with ${\bm Q} \parallel (1,1,0)$ is lower in the $\Gamma_4$ representation. So, the $\Gamma_4$ representation is found to be inconsistent with the experimental data in CeCoIn$_5$. 
}
  \label{fig.5}
\end{figure}

\begin{figure}[htbp]
 \begin{minipage}{0.5\hsize}
  \begin{center}
   \includegraphics[width=60mm]{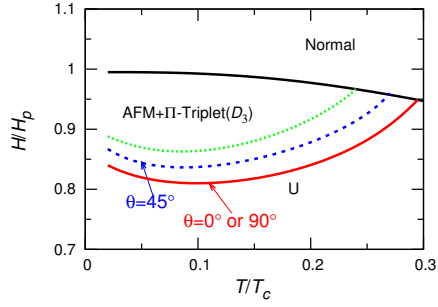}
\end{center}
 \end{minipage}
 \caption{Example of the HFLT phase in the $H$-$T$ phase diagram in the Pauli-limited case with the $\pi$-triplet order in the $\Gamma_3$ representation but with no FFLO order assumed. The parameter values $T_{c}/U=0.01622$ and $T_{c}/V_3=0.0033$ are used. The $\theta$ dependence of the free energy and of the stable SDW ${\bm Q}$-vector leads to the $\theta$ dependence of the field range of the HFLT phase, as indicated in the figure. The dotted curve is the transition line entering the HFLT phase in the case with no $\pi$-triplet order. 
}
  \label{fig.6}
\end{figure}

Rewriting eqs.(\ref{eq:Green_d3_11}) and (\ref{eq:Green_d3_2}), we have 
\begin{equation}
f^{(1, 1)}_{{\it \Pi}, m}
=
2T \sum^{}_{\varepsilon_{n}>0, {\bm p}, {\bm q}}
\biggl(
\frac{i D_{s}(\theta)|_\perp}{ c_{1}^{2}+d_{1}^{2} }
( c_{1} I ({\it \Delta}^{\ast}_{{\bm p}+{\bm Q}} - {\it \Delta}^{\ast}_{\bm p})
- d_{1} \varepsilon_{n} ({\it \Delta}^{\ast}_{{\bm p}+{\bm Q}} + {\it \Delta}^{\ast}_{\bm p})  )\nonumber \\ 
\times ( -m({\bm q}) {\it \Pi}^{(s)}_{-{\bm Q}}({\bm q}) 
+ m^{\ast}({\bm q})  {\it \Pi}^{(s)}_{\bm Q}({\bm q})   )
+ {\rm H.c.} \biggr) ,
\nonumber \\ 
\label{eq:14_d3_free11}
\end{equation}
\begin{eqnarray}
f^{(2)}_{\it \Pi}
&=& \sum_{\bm q} 
\frac{ |{\it \Pi}^{(s)}_{-{\bm Q}}({\bm q})|^{2} + |{\it \Pi}^{(s)}_{\bm Q}({\bm q})|^{2} }{V_{s}} 
- 2T \sum^{}_{\varepsilon_{n}>0, {\bm p}, {\bm q}}
\biggl( 
\frac{[D_{s}(\theta)|_\perp]^2  c_{1}}{ c_{1}^{2}+d_{1}^{2} } 
(\varepsilon_{n}^{2} + \varepsilon({\bm p})\varepsilon ({\bm p}+{\bm Q} ) +I^{2} ) \nonumber \\
&+& 
\frac{[D_{s}(\theta)|_\parallel]^2}{ e_{1}^{2}+g_{1}^{2} } 
( e_{1} (\varepsilon_{n}^{2} + \varepsilon({\bm p})\varepsilon ({\bm p}+{\bm Q}) -I^{2} )
+ g_{1} 2\varepsilon_{n} I)\biggr) 
\biggl( |{\it \Pi}^{(s)}_{-{\bm Q}}({\bm q})|^{2} + |{\it \Pi}^{(s)}_{\bm Q}({\bm q})|^{2} \biggr) \nonumber \\ 
&+&
2T \sum^{}_{\varepsilon_{n}>0, {\bm p}, {\bm q}}
\biggl( 
\biggl( \frac{[D_{s}(\theta)|_\perp]^2 c_{1}}{ c_{1}^{2}+d_{1}^{2} }
+
\frac{[D_{s}(\theta)|_\parallel]^2 e_{1}}{ e_{1}^{2}+g_{1}^{2} } \biggr)
{\it \Delta}^{\ast}_{\bm p} {\it \Delta}^{\ast}_{{\bm p}+{\bm Q}}
{\it \Pi}^{(s)}_{-{\bm Q}}({\bm q}) {\it \Pi}^{(s)}_{\bm Q}({\bm q}) 
+ {\rm H.c.} \biggr).
\nonumber \\ 
\label{eq:14_d3_free2}
\end{eqnarray}
Here, the coefficients $a_1$, $a_2$, $b_1$, $c_{1}$, and $d_{1}$ were defined in eqs.(\ref{eq:14_d3_free2}) and (\ref{eq:abcd}), and the coefficients $e_{1}$ and $g_{1}$ are 
\begin{eqnarray}
e_{1} &=& a_{1}a_{2} - b_{1}^{2}, \nonumber \\ 
g_{1} &=& b_{1} ( a_{2} + a_{1} ).
\end{eqnarray}

In the same manner as the case of the $\Gamma_1$ representation, the phase diagram is obtained like Fig.6. As in the $\Gamma_1$ representation, the presence of the $\pi$-triplet order of the $\Gamma_3$ representation also leads to a broadening of the HFLT phase. Further, as shown in Fig.5, the switching of ${\bm Q}$-vector upon sweeping the field direction of the type detected in the experiment \cite{Gerber} occurs. In contrast to the $\Gamma_1$ state, however, the field range in which the HFLT phase accompanied by the $\Gamma_3$ triplet pairing state is realized shows a remarkable angular dependence. Nevertheless, this triplet pairing state has {\it no} gap nodes along $(k, \pm k, 0.5)$ and thus, is believed to be different from the triplet pairing state suggested from the thermal conductivity experiment \cite{Kim}. In fact, by comparing Fig.6 with Fig.4. and noting the values of the coupling constants $V_1$ and $V_3$ used in the figures, the field range of the HFLT phase with $\Gamma_1$ is found to be broader than that with $\Gamma_3$ under the same value of $V_s$ ($s=1$ and $3$). This result suggests that the $\Gamma_1$ state is more stable than the $\Gamma_3$ one. 

\section{Switching of ${\bm Q}$-vector due to FFLO modulation}

In the preceding sections, it has been shown that the $\pi$-triplet pairing state expected to occur theoretically and suggested from the thermal conductivity data is insensitive to the in-plane direction of the applied magnetic field and thus, is not the origin of the switching of the SDW ${\bm Q}$-vector upon the in-plane rotation of the magnetic field. It has been shown elsewhere \cite{Hatake2015} that the FFLO spatial modulation parallel to the magnetic field, which is believed to be present in the HFLT phase on the basis of various experimental facts \cite{Kumagai1,Tokiwa}, can become the origin of the switching of the ${\bm Q}$-vector. That is, in the notation of Fig.3, when the in-plane field ${\bm H}$ is oriented to any direction between [110] and [100] so that $0 \leq \theta \leq \pi/4$, the SDW ${\bm Q}$ is parallel to $(k, -k, 0.5)$, while the SDW ${\bm Q}$ becomes parallel to $(k, k, 0.5)$ when $\pi/4 \leq \theta \leq \pi/2$. In this section, the switching of the SDW ${\bm Q}$-vector is revisited and will be explained within the Pauli-limited FFLO theory neglecting the presence of the vortices, because inclusion of the $\pi$-triplet SC order to be done in the next section is performed for convenience in the Pauli limit. 

For the purpose of the present section mentioned above, we need to take account of a spatial modulation of the $d$-wave SC order parameter, while the presence of the $\pi$-triplet order will be neglected. Then, in this section we will use eq.(10) with no ${\cal H}_{\rm TS}$. 

Following previous works and using the expressions of the $d$-wave SC and SDW order parameters in the FFLO phase with a spatial modulation parallel to the magnetic field 
\begin{eqnarray}
{\it \Delta} ({\bm R}) &=& {\it \Delta} \sqrt{2} \cos({\bm q}_{\rm LO} \cdot {\bm R}), \nonumber \\
m ({\bm q}, {\bm R}) &=& |m({\bm q})| \sqrt{2} \cos({\bm q}_{\rm LO} \cdot {\bm R} + \delta_m), 
\end{eqnarray}
we will derive the free energy including the gradient terms here in the form
\begin{eqnarray}
f &=& f_{\it \Delta}(q_{\rm LO})+f_{m}(q_{\rm LO}) \nonumber\\
&=& f_{{\it \Delta}, (0)}+f_{{\it \Delta}, (2)}+f_{{\it \Delta}, (4)}+ \cdots +f^{(2)}_{m}(q_{\rm LO})+f^{(4)}_{m}(q_{\rm LO}) + \cdots,
\label{eq:free}
\end{eqnarray}
where the relative phase $\delta_m$ will be determined by minimizing the free energy (see also the following figures). The first three terms consist only of the $d$-wave SC order parameter ${\it \Delta}({\bm R})$ with FFLO spatial modulations. Using the expression of the SC 
free energy \cite{Gert}
\begin{eqnarray}
f_{\it \Delta}(q_{\rm LO})
=\biggl\langle 
 \frac{|{\it \Delta}({\bm R})|^{2}}{|g|} 
+ \frac{T}{2}
\sum^{\infty}_{\varepsilon_{n}=-\infty} \sum^{}_{{\bm p},\sigma} \int_{\varepsilon_{n}}^{\infty s_{\epsilon}}
\!d\omega\  \mathop{\mathrm{Tr}}
\biggl[ i \sigma_{z} {\hat G}^{(\sigma)}_{\omega}({\bm p},{\bm R}) \biggr]
\biggr\rangle_{\bm R} 
\label{eq:fdel}
\end{eqnarray}
with $\langle \,\,\, \rangle_{\bm R}$ implying the average over the center of mass coordinate ${\bm R}$ of the Cooper pair and the results on the gradient expansion for the Green's function 
${\hat G}^{(\sigma)}=\hat{G}^{(\sigma)}_{(0)}+{\hat G}^{(\sigma)}_{(1)}+\hat{G}^{(\sigma)}_{(2)}
+{\hat G}^{(\sigma)}_{(3)}+\hat{G}^{(\sigma)}_{(4)}
+\cdots, $ 
where 
\begin{equation}
{\hat G}^{(\sigma)}_{\varepsilon_n, \, (m)}({\bm p},{\bm R}) = - i {\hat G}^{(\sigma)}_{(0)} \biggl( {\bm v}_{\bm p}\cdot\nabla_{\bm R} 
{\hat G}^{(\sigma)}_{\varepsilon_n, \, (m-1)} \biggr) 
,\label{eq:G4}  
\end{equation}
($m=1$, $2$, $3$, or $4$), 
$f_{{\it \Delta}, (0)}$, $f_{{\it \Delta}, (2)}$, and $f_{{\it \Delta}, (4)}$ are expressed \cite{Hosoya} as 
\begin{eqnarray}
f_{{\it \Delta},(0)} &=& \biggl\langle \frac{|{\it \Delta}({\bm R})|^2}{|g|} - T \sum_{\varepsilon_n>0} \sum_{\bm p}
	\ln \biggl[\frac{(\varepsilon_n^2+[\varepsilon({\bm p})]^2+|{\it \Delta}_{\bm p}({\bm R})|^2-I^2)^2+4 \varepsilon_n^2 I^2}{(\varepsilon_n^2+[\varepsilon({\bm p})]^2-I^2)^2+4 \varepsilon_n^2 I^2} \biggr] \biggr\rangle_{\bm R}, \nonumber \\ 
f_{{\it \Delta}, (2)} &=& \biggl\langle T \sum_{\varepsilon_n>0} \sum_{\bm p} \biggl[\frac{a_1^2-b_1^2}{(a_1^2+b_1^2)^2} |{\bm v}_{\bm k}\cdot \nabla {\it \Delta}_{\bm p}({\bm R})|^2 \nonumber \\
&+& \frac{2}{3} \frac{(2[\varepsilon({\bm p})]^2-\varepsilon_n^2+I^2-|{\it \Delta}_{\bm p}({\bm R})|^2) 
(a_1^4-6a_1^2b_1^2+b_1^4) - 4 a_1 b_1^2 (a_1^2-b_1^2) }{(a_1^2+b_1^2)^4} ({\bm v}_{\bm p}\cdot\nabla |{\it \Delta}_{\bm p}({\bm R})|^2)^2 \biggr] \biggr\rangle_{\bm R}, 
\nonumber \\
f_{{\it \Delta}, (4)} &\simeq& \biggl\langle T \sum_{\varepsilon_n>0} \sum_{\bm p} 
\biggl[ \frac{2}{3} \frac{(2 [\varepsilon({\bm p})]^2-\varepsilon_n^2+I^2-|{\it \Delta}_{\bm p}({\bm R})|^2) 
(a_1^4-6a_1^2b_1^2+b_1^4) - 4 a_1 b_1^2 (a_1^2 - b_1^2)}{(a_1^2+b_1^2)^4}|({\bm v}_{\bm p}\cdot \nabla)^2 {\it \Delta}_{\bm p}({\bm R})|^2 \biggr] \biggr\rangle_{\bm R}, 
\label{FSCgrad}
\end{eqnarray}
where $\Delta_{\bm p}({\bm R}) = \Delta({\bm R}) w_{\bm p}$. 

Next, the free energy term $f_{\rm m}$ associated with the SDW order parameter $m$ in eq.(\ref{eq:free}) will be derived in the form 
i.e., 
\begin{eqnarray}
f_{m} &=& f^{(2)}_{m}(q_{\rm LO})+f^{(4)}_{m}(q_{\rm LO}) + \cdots. \nonumber\\
&=& f^{(2, 0)}_{m, q_{\rm LO}}+f^{(2, 2)}_{m, q_{\rm LO}}+ \cdots. + f^{(4, 0)}_{m, q_{\rm LO}}+ \cdots
\label{eq:F_pertubation}
\end{eqnarray}
expressed as the GL expansion about both of $m$ and the FFLO wavenumber $q_{\rm LO}$.  Here, $|q_{\rm LO}|$ is the order parameter of the FFLO state. It is found that, as is shown in Fig.9 below, $|q_{\rm LO}|$ in equilibrium is proportional to $\sqrt{H-H_{\rm LO}(T)}$ when the field-induced transition entering the FFLO state is a second order transition on $H_{\rm LO}(T)$. This behavior has been found in the NMR data of Ref.2 by assuming $H_{\rm LO}$ to coincide with $H^*(T)$ defined in sec.III (see Fig.5 (b) in Ref.2). 

First, the O($m^2$) term 
\begin{eqnarray}
f^{(2)}_{m}(q_{\rm LO}) = \sum_{\bm q} \biggl\langle 
\biggl[
\frac{1}{U} 
+ \frac{T}{2}
\sum^{}_{{\bm p},\sigma, \varepsilon_{n}} 
\sum^{}_{s_{1}, s_{2} = \pm 1} 
{\rm Tr}
\biggl(
{\hat G}^{(\sigma)}_{\varepsilon_n}({\bm p} + {\bm Q}_0 + s_{1} {\bm q} + s_{2} {\bm q}_{\rm LO},{\bm R})
{\hat G}^{(-\sigma)}_{\varepsilon_n}({\bm p},{\bm R})
\biggr)
\biggr]
|m({\bm q}, {\bm R})|^{2} \biggr\rangle_{\bm R} 
\end{eqnarray}
will be rewritten in the form expanded w.r.t. ${q}_{\rm LO}$. Using 
\begin{eqnarray}
\varepsilon({\bm p}+{\bm Q}+{\bm q}_{\rm LO}) & \simeq &
\varepsilon({\bm p}+{\bm Q}) 
+ {\bm q}_{\rm LO} \cdot {\bm v}_{{\bm p}+{\bm Q}} + \frac{1}{2}({\bm q}_{\rm LO} \cdot \nabla_{\bm p})^2 \varepsilon({\bm p}+{\bm Q}),
\nonumber\\
\nabla_{\bm p} &=& (\frac{\partial}{\partial p_x}, \frac{\partial}{\partial p_y},  \frac{\partial}{\partial p_z})
\nonumber\\
{\it \Delta}_{{\bm p}+{\bm Q}+{\bm q}_{\rm LO}}({\bm R})
& \simeq & 
{\it \Delta}_{{\bm p}+{\bm Q}}({\bm R})
+ {\bm q}_{\rm LO} \cdot ( \nabla_{\bm p} {\it \Delta}_{{\bm p}+{\bm Q}}({\bm R}) )
+ \frac{1}{2} ({\bm q}_{\rm LO} \cdot  \nabla_{\bm p})^2 {\it \Delta}_{{\bm p}+{\bm Q}}({\bm R}), 
\label{eq:expansion}
\end{eqnarray}
where 
\begin{eqnarray}
{\hat G}^{(\sigma)}_{\varepsilon_n, \, (0)}({\bm p}+{\bm Q}_{0}+s_{1} {\bm q}+s_{2} {\bm q}_{\rm LO}, {\bm R}) &=& 
{\hat G}^{(\sigma)}_{\varepsilon_n, \, (0, 0)}({\bm p}+{\bm Q}_{0}+s_{1} {\bm q} ,{\bm R})
+ {\hat G}^{(\sigma)}_{\varepsilon_n, \, (0, 2)}({\bm p}+{\bm Q}_{0}+s_{1} {\bm q} ,{\bm R})+\cdots,
\end{eqnarray}
the second term on the second row of eq.(42) is expressed as 
\begin{eqnarray}
f^{(2, 2)}_{m, q_{\rm LO}} &=& \biggl\langle
\frac{T}{2}
\sum^{}_{{\bm p}, {\bm q}, \sigma, \varepsilon_{n}} 
\sum^{}_{s_{1} = \pm 1} 
{\rm Tr}
\biggl[
{\hat G}^{(\sigma)}_{\varepsilon_n, \, (2)}({\bm p}+{\bm Q}_{0}+s_{1} {\bm q}, {\bm R})
{\hat G}^{(-\sigma)}_{\varepsilon_n, \, (0)}({\bm p}, {\bm R})
+
{\hat G}^{(\sigma)}_{\varepsilon_n, \, (1)}({\bm p}+{\bm Q}_{0}+s_{1} {\bm q}, {\bm R})
{\hat G}^{(-\sigma)}_{\varepsilon_n, \, (1)}({\bm p}, {\bm R})\nonumber\\
&+&
{\hat G}^{(\sigma)}_{\varepsilon_n, \, (0,0)}({\bm p}+{\bm Q}_{0}+s_{1} {\bm q}, {\bm R})
{\hat G}^{(-\sigma)}_{\varepsilon_n, \, (2)}({\bm p}, {\bm R})
+
{\hat G}^{(\sigma)}_{\varepsilon_n, \, (0, 2)}({\bm p}+{\bm Q}_{0}+s_{1} {\bm q}, {\bm R})
{\hat G}^{(-\sigma)}_{\varepsilon_n, \, (0)}({\bm p}, {\bm R})
\biggr]
|m({\bm q}, {\bm R})|^{2} \biggr\rangle_{\bm R},
\label{eq:m2q2}
\end{eqnarray}
while 
$f^{(2, 0)}_{m, q_{\rm LO}}$ and $f^{(4, 0)}_{m, q_{\rm LO}}$ are given by replacing ${\it \Delta}$ in eqs.(\ref{eq:free_m2}) and (\ref{eq:free_m4}) with 
${\it \Delta}({\bm R})$ and taking their space average over ${\bm R}$. 

The procedure for rewriting the cross term $f^{(2, 2)}_{m, q_{\rm LO}}$ is involved and will be explained in Appendix. The resulting $f^{(2, 2)}_{m, q_{\rm LO}}$ depends on the relative orientation between ${\bm q}_{\rm LO}$, which is parallel to the magnetic field, and the crystal axis reflected in the dispersion relation $\varepsilon({\bm p})$, that is, on the angle $\theta$ defined in Fig.3. Its calculated result is shown in Fig.7. Since no information on the SDW ${\bm Q}$-vector is included in $f_{\it \Delta}$ which primarily determines the value of $q_{\rm LO}$, the stable ${\bm Q}$-direction at each $\theta$ is determined only from Fig.7 as far as $|m|$ value is so small that the GL expansion on $|m|$ is justified. 

\begin{figure}[htbp]
  \begin{center}
   \includegraphics[width=60mm]{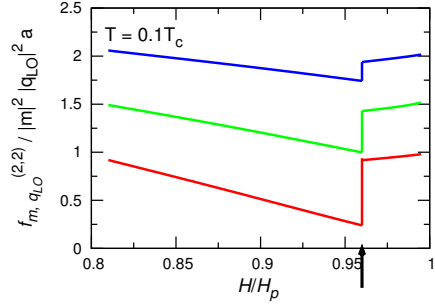}
  \hspace{1.2cm} 
\end{center}
 \caption{Field dependences of $f^{(2, 2)}_{m, q_{\rm LO}}$ at the temperature $T=0.1 T_c$ when $\theta$ (see Fig.3) is zero (upper blue curve), $\pi/4$ (middle green one), and $\pi/2$ (bottom red one). The parameter value $T_c/U=0.01597$ was used. At the field $H/H_{P}=0.96$ indicated by the arrow, the configuration of $m({\bm R})$ relative to the FFLO modulation of the $d$-wave SC order parameter ${\it \Delta} ({\bm R})= {\it \Delta} \sqrt{2} \cos({\bm q}_{\rm LO} \cdot {\bm R})$ shows such a structural transition \cite{Hosoya,Hatake1,Tayama} that, in $H/H_{P}<0.96$, $\delta_m=\pi/2$, while $\delta_m=0$ in $H/H_{P}>0.96$ (see eq.(37) and Fig.11 (a)). 
}
 \label{fig.7}
\end{figure}

According to the previous work \cite{Hatake1}, the incommensurate part ${\bm q}={\bm Q} - {\bm Q}_0$ of the SDW wavevector tends to become parallel to ${\bm Q}_0$. Hence, ${\bm Q}$ favors one of the gap node directions of ${\it \Delta}$. Further, according to Fig.7, ${\bm q}$ favors a more separated direction from the in-plane magnetic field to which the direction of the FFLO modulation of ${\it \Delta}$ is parallel. Therefore, the in-plane component of ${\bm Q}$ is parallel to [1,-1,0] when $0 \leq \theta \leq \pi/4$, while it is directed along [1,1,0] when $\pi/4 \leq \theta \leq \pi/2$. This is the explanation on the experimental observation in Ref.15 based on the FFLO theory. 

\begin{figure}[htbp]
  \begin{center}
   \includegraphics[width=60mm]{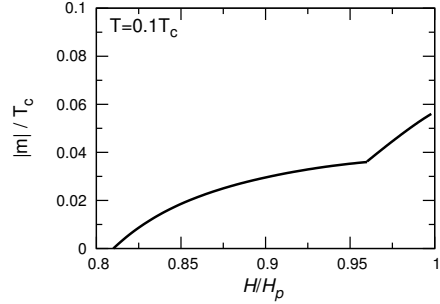}
  \hspace{1.2cm} 
\end{center}
 \caption{Field dependence of $|m|/T_c$ at $T= 0.1 T_c$. In higher fields ($H > H_P$), $|m|$ vanishes discontinuously. The used parameter values are the same as those in Figs.7 and 11(a).
}
 \label{fig.8}
\end{figure}

\begin{figure}[htbp]
  \begin{center}
   \includegraphics[width=60mm]{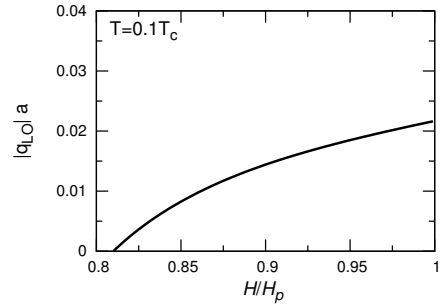}
  \hspace{1.2cm} 
\end{center}
 \caption{Field dependence of $|q_{\rm LO}|$ at $T= 0.1 T_c$. The used parameter values are the same as those in Figs.7 and 11(a). Here, the parameter $a$ normalizing $|q_{\rm LO}|$ is the lattice constant in the basal plane. Note that, in spite of the absence of impurities in the present model, the $|q_{\rm LO}|$-value is notably small. 
}
 \label{fig.9}
\end{figure}

The phase diagram following from the analysis in this section will be shown later (see Fig.11 (a)). 
Strictly speaking, the phase diagram also depends upon $\theta$. However, as far as the FFLO wavenumber is so small that the coupling between the SDW and FFLO orderings can be regarded as being weak, such a $\theta$ dependence of the phase diagram is expected to be negligibly small. 

The structural transition at $H=0.96 H_P$ indicated in Fig.7 should be reflected in some quantities. In the field dependence of the magnitude $|m|$ of the 
SDW order parameter shown in Fig.8, the structure transition is reflected as a visible upturn of the $|m(H)|$ curve. It should be stressed that such an upturn of the field dependence can be seen in the internal field, corresponding to $|m|$, taken from NMR data in Ref.24 (see a feature around 10.8(T) in Fig.2 of Ref.24). Note that the internal field shown there \cite{KMF} has an upwardly curved field variation in higher fields, although, conventionally, the magnitude of the order parameter tends to saturate far from the phase boundary. Such a remarkable anomaly at about 11 (T) has also been seen previously in the data associated with the magnetization \cite{Tayama}. 

In Fig.9, the field variation of another order parameter $q_{\rm LO}$ characterizing the HFLT phase is shown. Since $q_{\rm LO}$ is inversely proportional to the distance between the neighboring FFLO nodal planes, the field dependence of $|q_{\rm LO}|$ shows that of the number of excess quasi particles occurring in the FFLO state with the one-dimensional spatial modulation parallel to the field. The fact \cite{Kumagai1} that the excess DOS in the HFLT phase detected experimentally is proportional to $\sqrt{H-H^*}$ near the $H^*(T)$-line strongly suggests the presence of the FFLO modulation in the HFLT phase. A further reduction of $|q_{\rm LO}|$ due to inclusion of impurities was argued in Ref.4 to result in the detected suppression \cite{Tokiwa} of the ordering itself forming the HFLT phase.

\section{HFLT phase with $\pi$-triplet pairing order}

In the preceding sections, we have shown that the switching of the SDW ${\bm Q}$-vector upon rotaing the magnetic field in the basal plane is explained by the presence of the FFLO spatial modulation parallel to the magnetic field, and that the recent thermal conductivity data indicate the presence of the $\pi$-triplet order ${\bm D}_1$ in the $\Gamma_1$ representation. In this section, we examine how the presence of the ${\bm D}_1$ order affects the phase boundaries associated with the HFLT phase. 

For this purpose, we only have to take account of the three novel orders, FFLO, SDW, and the $\pi$-triplet ones, altogether. Since, in the present theory, the $\pi$-triplet order is the secondary order induced by the SDW order which the FFLO spatial modulation enhances \cite{IHA,Hatake1}, any direct coupling of the $\pi$-triplet order to the FFLO order may be neglected. Under this assumption, the mean field analysis roughly explained in sec.III can straightforwardly be performed, because one has only, as done in sec.III, to minimize the free energy w.r.t. the $\pi$-triplet order. To perform this in the lowest order in $q_{\rm LO}$, the free energy terms ${\overline f}_m$ and ${\overline f}_{\it \Pi}$ which take the place of $f_m$ and $f_{\it \Pi}$ in eq.(22), respectively, will be considered. Here, ${\overline f}_m$ (${\overline f}_{\it \Pi}$) is the average of $f_m({\bm R})$ ($f_{\it \Pi}({\bm R})$) over ${\bm R}$, where $f_m({\bm R})$ ($f_{\it \Pi}({\bm R})$) is given by $f_m$ ($f_{\it \Pi}$) in eq.(22) with the order parameters $m$ and ${\it \Delta}$ replaced simply by $m({\bm R})$ and ${\it \Delta}({\bm R})$, respectively. Then, minimization over the ${\bm R}$-dependent $\pi$-triplet order parameter ${\it \Pi}^{(1)}_{\bm Q}({\bm q}; {\bm R})$ leads to 

\begin{eqnarray}
{\it \Pi}^{(1)}_{-{\bm Q}}({\bm q},{\bm R}) &=& 
\frac{-2T \sum_{\varepsilon_{n}>0, {\bm p}} 
(\varepsilon({\bm p}) 
{\it \Delta}^{\ast}_{{\bm p}+{\bm Q}}({\bm R}) 
- \varepsilon({\bm p}+{\bm Q}){\it \Delta}^{\ast}_{\bm p}({\bm R}) )
w_{\bm p} c_{1}/( c_{1}^{2}+d_{1}^{2}) } 
{ V_1^{-1} 
- 2T \sum_{\varepsilon_{n}>0, {\bm p}} 
( \varepsilon_{n}^{2}+I^{2} + \varepsilon({\bm p})\varepsilon({\bm p}+{\bm Q}) 
+ {\it \Delta}_{\bm p}({\bm R}) {\it \Delta}^{\ast}_{{\bm p}+{\bm Q}}({\bm R}) )
w_{\bm p}^{2} c_{1}/(c_{1}^{2}+d_{1}^{2}) }  
m({\bm q},{\bm R}).
\nonumber \\
\label{eq:pi_min_LO_Pi}
\end{eqnarray}

We note that, as far as the ${\bm R}$ dependences are concerned, this expression can simply be written as 
\begin{equation}
{\it \Pi}^{(1)}_{-{\bm Q}}({\bm q},{\bm R})
= C({\bm R}) {\it \Delta}^{\ast}({\bm R}) m({\bm q},{\bm R}),
\end{equation}
where the coefficient $C({\bm R})$ depends on $|{\it \Delta}({\bm R})|^2$ and includes a spacial dependence due to the higher order terms of the GL expansion in ${\it \Delta}({\bm R})$. However, the ${\bm R}$ dependence of $C({\bm R})$ is a quantitatively weak effect so that $C$ may be regarded as a constant. Thus, in the low field region of the HFLT phase where the SDW order parameter has the out-of-phase configuration, $\delta_m=\pi/2$, with ${\it \Delta}({\bm R})$, ${\it \Pi}^{(1)}$ behaves like ${\rm sin}(2 {\bm q}_{\rm LO}\cdot{\bm R})$, while it takes the form $1 -  {\rm cos}(2 {\bm q}_{\rm LO}\cdot{\bm R})$ in higher fields. The resulting structural transition line \cite{Hosoya,Hatake1,Tayama,KMF} separating the two configuration, sketched in Fig.10, from each other is expressed by the thin dotted line in Fig.11. 

By substituting eq.(47) into ${\overline f}_{\it \Pi}$, an additional term to ${\overline f}_m$ is obtained. The resulting free energy composed only of ${\it \Delta}$ and $m$ leads to the phase diagram shown in Fig.11 (b). For comparison, the corresponding result with no $\pi$-triplet order included is also presented in Fig.11 (a). 

\begin{figure}[t]
\scalebox{0.6}[0.6]{\includegraphics{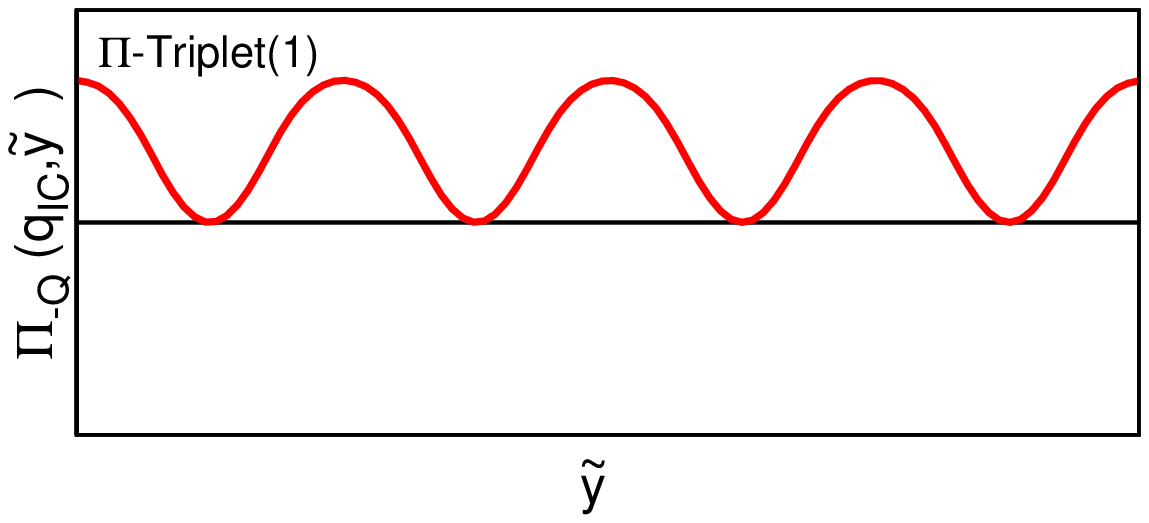}}
\scalebox{0.6}[0.6]{\includegraphics{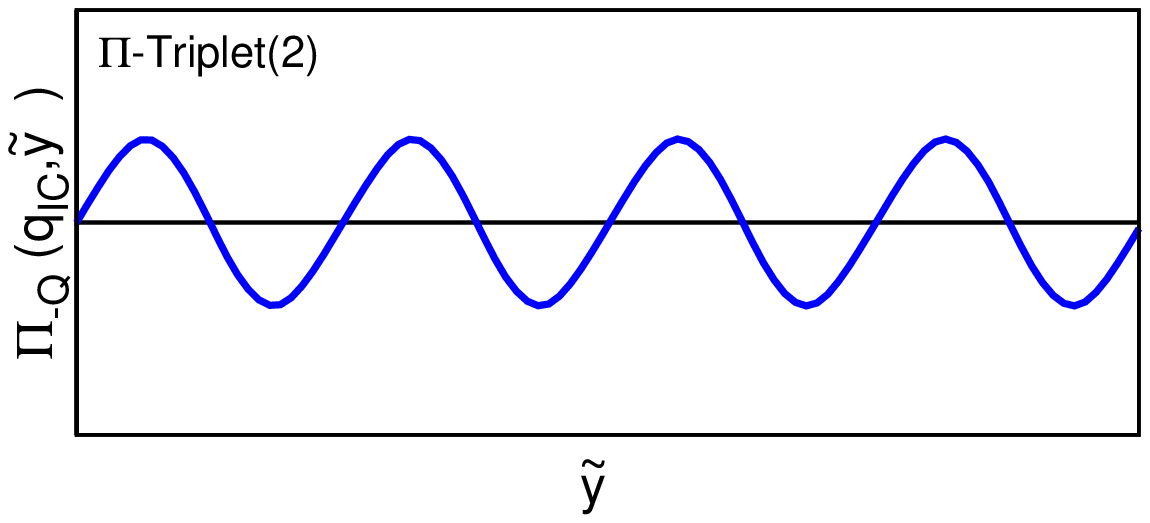}}
\caption{Spatial modulation of ${\it \Pi}^{(1)}_{-{\bm Q}}({\bm q},{\bm R})$  in the ${\tilde y}$ direction, parallel to the in-plane field direction, in the cases where the SDW order parameter $m({\bm R})$ shows the (1) in-phase ($\delta_m=0$) and (2) out-of-phase ($\delta_m=\pi/2$) modulation with the FFLO variation ${\it \Delta}({\bm R})= {\it \Delta} \sqrt{2} \cos({\bm q}_{\rm LO} \cdot {\bm R})$ of the $d$-wave SC order parameter. 
}
\label{fig.10}
\end{figure}

It can be seen from the figures that inclusion of the $\pi$-triplet order leads to diminishing of the pure FFLO region with no SDW order and makes the concave form of the second order transition curve on entering the SDW phase a convex one which is consistent with the experimental result \cite{Bianchi,Kenzel,Kim}. We argue that this change of the high field phase diagram due to inclusion of the $\pi$-triplet order will be an improvement on the theoretical description of the HFLT phase of CeCoIn$_5$. Another reduction of the pure FFLO region can be expected in higher fields, i.e., at higher temperatures, by including the quantum SDW critical fluctuation \cite{Hatake1,RI76}. 

\begin{figure}[htbp]
 \begin{minipage}{0.45\hsize}
  \begin{center}
   \includegraphics[width=60mm]{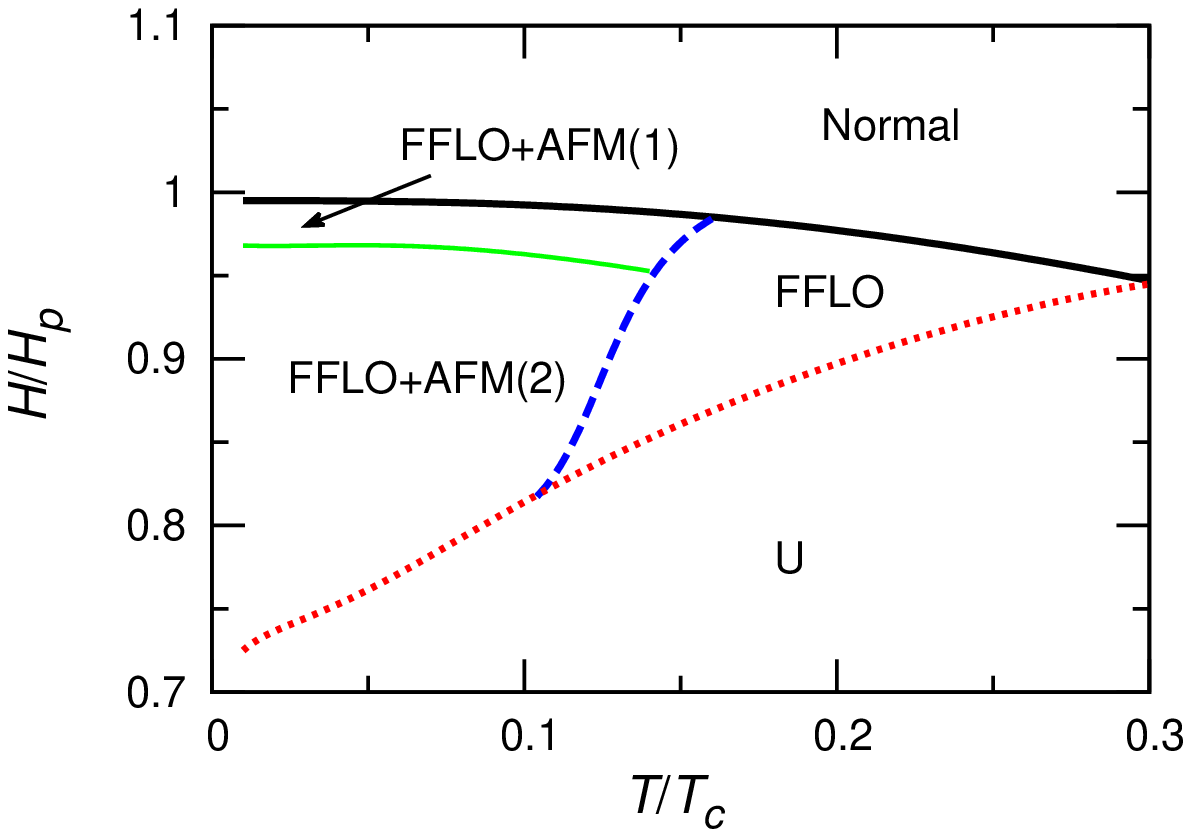}
  \hspace{1.6cm} (a)
\end{center}
 \end{minipage}
 \begin{minipage}{0.45\hsize}
  \begin{center}
  \includegraphics[width=60mm]{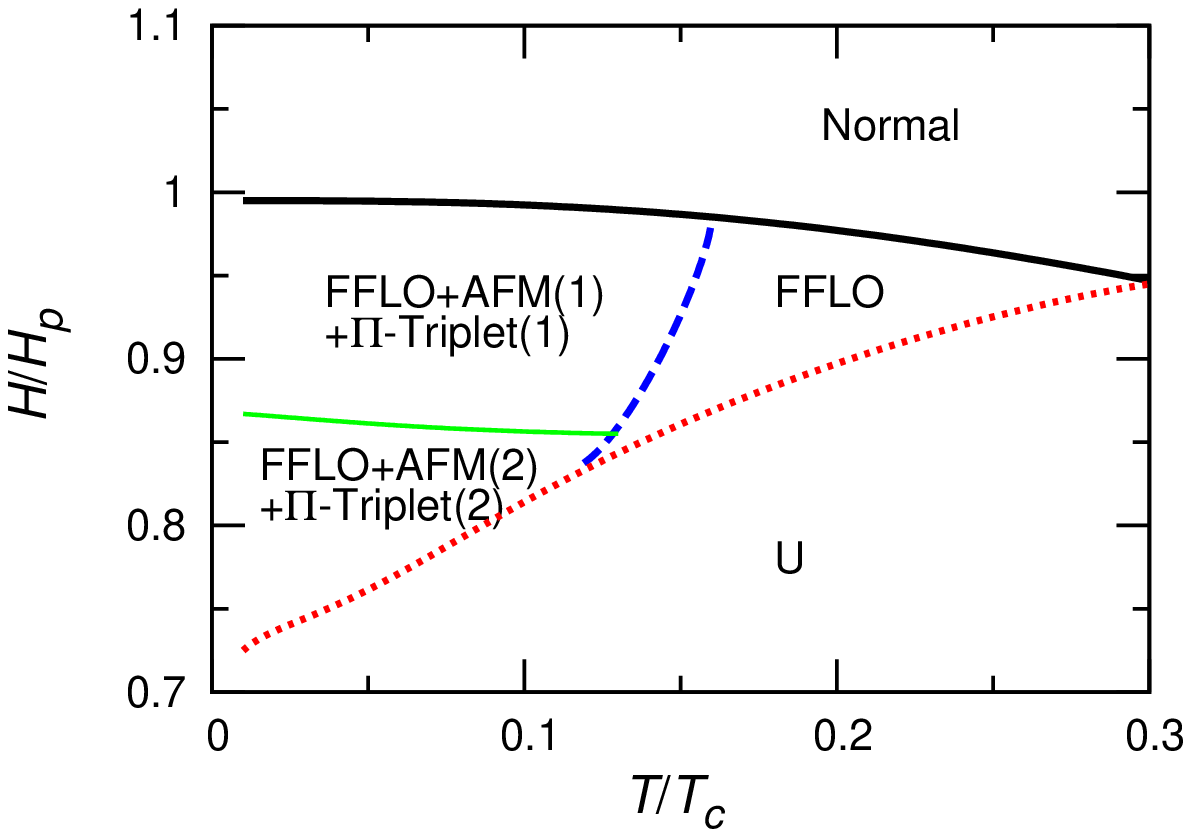}
 \hspace{1.6cm} (b)
 \end{center}
 \end{minipage}
 \caption{$H$-$T$ phase diagram (a) obtained using the parameter value $T_{c}/U=0.01597$ with no $\pi$-triplet order and (b) obtained using the parameters $T_c/U=0.01654$ and $T_c/V_1=0.01125$ and incorporating the $\pi$-triplet order ${\bm D}_1$. The parameter values were chosen so that the FFLO transition (dotted black) curve remains unchanged irrespective of the presence or absence of the $\pi$-triplet order. In these high fields, the $H_{c2}$-transition on the thick solid (black) curve is of first order in the mean field approximation. The second order transition line $H^*(T)$ (thick dashed blue curve) at which $|m|$ begins to become nonzero is shifted to higher temperatures and became convex by including the nonzero $|{\bm D}_1|$ though it was concave with no $\pi$-triplet order. The thin solid (green) line which separates the in-phase configuration of the SDW order parameter $m$ from its out-of-phase one is shifted to lower fields by including the $\pi$-triplet order. 
}
  \label{fig.11}
\end{figure}

\section{Summary and Discussion}

In the present work, we use the Pauli-limited model neglecting the presence of the vortices and have extended the theory based on the strong paramagnetic pair-breaking (PPB) of the HFLT phase of the $d$-wave superconductor CeCoIn$_5$ to the case including the $\pi$-triplet SC pairing order which may accompany the PPB-induced SDW order. It has been shown that the switching of the SDW ${\bm Q}$-vector upon rotating the magnetic field ${\bm H}$ parallel to the basal plane cannot be explained based only on the presence of the stable $\pi$-triplet order of the type suggested from the recent thermal conductivity measurement \cite{Kim}, and that, as pointed out previously \cite{Hatake2015}, the presence of the FFLO spatial modulation parallel to ${\bm H}$ of the $d$-wave SC order parameter leads to the switching of the ${\bm Q}$-vector. Further, due to the presence of the $\pi$-triplet order, further agreement on the phase diagram between the experimental data and the result of the present theory based on the strong 
PPB have been reached. 

In the present theory, the FFLO state with {\it no} SDW order inevitably appears at higher temperatures although, as suggested in sec.V, there are mechanisms leading to a shrinkage of this region. In CeCoIn$_5$, the appearance of the SDW order seems to occur at almost the same field as that of the FFLO modulation \cite{Kumagai1} at least at low enough temperatures. However, a different NMR experiment seems to have suggested the presence of the FFLO order with no SDW order \cite{KMF} at lower fields and higher temperatures. As argued in Ref.18, the presence of the FFLO state with no SDW order should be seen more clearly in experiments performed under a magnetic field tilted from the $a$-$b$ plane. 

Regarding the resulting $\pi$-triplet order, we need to give some comments associated with the thermal conductivity experiment \cite{Kim}. The thermal conductivity sees an additional DOS due to the Doppler shift of the quasiparticles \cite{Volovik}. This Doppler shift is given in the present context by the scalar product between the QP velocity ${\bm v}_{\bm p}$ and the SDW ${\bm Q}$-vector under the definition of the linearized SC gap function which is, in the case of the $\pi$-triplet order of our interest, ${\bm d}_1({\bm p})$ defined in sec.I. Or, in the tight-binding model, ${\bm d}_1$ for ${\bm Q}=(k, \pm k, 0.5)$ is replaced by ${\tilde {\bm d}}_1({\bm p}) = (0,0,{\rm sin}p_x \mp {\rm sin}p_y) = {\bm D}_1({\bm p} - {\bm Q}/2)$. That is, although one feels as if the switching of the SDW ${\bm Q}$-vector between the two directions $(q,\pm q, 0.5)$ induces that between the two triplet order parameters of ${\tilde {\bm d}}_1$ with different gap nodes, such a switching of the triplet order cannot be seen in the alternative representation ${\bm D}_1({\bm p})$ of the same triplet order (see sec.I). 

It has been argued elsewhere \cite{Mineev} that the switching of the SDW ${\bm Q}$-vector on rotating the in-plane magnetic field can be explained just by incorporating effects of a spin-orbit coupling on the band structure. However, it is unclear whether this approach leads to a quantitativey reasonable effect as far as the field-induced vortices are neglected, since it is known \cite{Machida} that the presence of the vortices, neglected in the work \cite{Mineev}, favors the SDW ${\bm Q}$-vector parallel to the magnetic field in contrast to the observation \cite{Gerber}. It should be stressed that, as mentioned in the preceding sections, there are experimental facts consistent with the presence of a spatial modulation parallel to the magnetic field in the HFLT phase \cite{Kumagai1,Tokiwa,Tayama,KMF}.

\begin{acknowledgments}
The present research of R.I. was supported by Grant-in-Aid for Scientific Research [No.16K05444] from MEXT, Japan. 
\end{acknowledgments}


\begin{thebibliography}{99}
\bibitem{Bianchi} A. Bianchi, R. Movshovich, C. Capan, P. G. Pagliuso, and J. L. Sarrao, Phys. Rev. Lett. {\bf 91}, 187004 (2003). 
\bibitem{Kumagai1} K. Kumagai, H. Shishido, T. Shibauchi, and Y. Matsuda, Phys. Rev. Lett. {\bf 106}, 137004 (2011). 
\bibitem{Tokiwa} Y. Tokiwa, R. Movshovich, F. Ronning, E.D. Bauer, P. Papin, A.D. Bianchi, J.F. Rauscher, S.M. Kauzlarich, and Z. Fisk, Phys. Rev. Lett. {\bf 101}, 037001 (2008); Y. Tokiwa, R. Movshovich, F. Ronning, E.D. Bauer, A.D. Bianchi, Z. Fisk, and J.D. Thompson, Phys. Rev. B {\bf 82}, 220502 (2010). 
\bibitem{RIfragile} R. Ikeda, Phys. Rev. B {\bf 81}, 060510(R) (2010).  
\bibitem{Kenzel} M. Kenzelmann, T. Strassle, C. Niedermayer, M. Sigrist, B. Padmanabhan, M. Zolliker, A.D. Bianchi, R. Movshovich, E.D. Bauer, J.L. Sarrao, and J.D. Thompson, Science {\bf 321}, 1652 (2008); 
M. Kenzelmann, S. Gerber, N. Egetenmeyer, J.L. Gavilano, T. Strassle, A.D. Bianchi, E. Ressouche, R. Movshovich, E.D. Bauer, J.L. Sarrao, and J.D. Thompson, Phys. Rev. Lett. {\bf 104}, 127001 (2010). 
\bibitem{SIKEDA} S. Ikeda,  H. Shishido,  M. Nakashima,  R. Settai,  D. Aoki,  Y. Haga,  H. Harima,  Y. Aoki,  T. Namiki,  H. Sato, and Y. Onuki, J. Phys. Soc. Jpn. {\bf 70}, 2248 (2001).  
\bibitem{Izawa} K. Izawa, H. Yamaguchi, Yuji Matsuda, H. Shishido, R. Settai, and Y. Onuki, Phys. Rev. Lett. {\bf 87}, 057002 (2001). 
\bibitem{Ada} H. Adachi and R. Ikeda, Phys. Rev. B {\bf 68}, 184510 (2003). 
\bibitem{Yanase} See also Y. Yanase, J. Phys. Soc. Jpn. {\bf 77}, 063705 
(2008). 
\bibitem{IHA} R. Ikeda, Y. Hatakeyama, and K. Aoyama, Phys. Rev. B {\bf 82}, 060510(R) (2010). 
\bibitem{other} See also V. P. Michal and V. P. Mineev, Phys. Rev. B {\bf 84}, 052508 (2011); Y. Kato, C. D. Batista, and I. Vekhter, Phys. Rev. Lett. {\bf 107}, 096401 (2011); B. Rosemeyer and A. Vorontsov, Phys. Rev. B {\bf 89}, 220501 (2014). 
\bibitem{Hatakefinal} Y. Hatakeyama and R. Ikeda, Phys. Rev. B {\bf 93}, 104503  (2016). 
\bibitem{Agter} D. Agterberg, M. Sigrist, and H. Tsunetsugu, Phys. Rev. Lett. {\bf 102}, 207004 (2009). 
\bibitem{Aperis} A. Aperis, G. Varelogiannis, P. B. Littlewood, Phys. Rev. Lett. {\bf 104}, 216403 (2010). 
\bibitem{Gerber} S. Gerber, M. Bartkowiak, J. L. Gavilano, E. Ressouche, N. Egetenmeyer, C. Niedermayer, A. D. Bianchi, R. Movshovich, E. D. Bauer, J. D. Thompson, and M. Kenzelmann, Nature Physics {\bf 10}, 126 (2014).  
\bibitem{Hatake2015} Y. Hatakeyama and R. Ikeda, Phys. Rev. B {\bf 91}, 094504 (2015). 
\bibitem{Kim} D. Y. Kim, S-Z. Lin, F. Weickert, M. Kenzelmann, E. D. Bauer, F. Ronning, J. D. Thompson, and R. Movshovich, Phys. Rev. X {\bf 6}, 041059 
(2016). 
\bibitem{Hosoya} K. Hosoya and R. Ikeda, Phys. Rev. B {\bf 88}, 094513 (2013). 
\bibitem{RE} M. R. Eskildsen, private communication. 
\bibitem{Hatake1} Y. Hatakeyama and R. Ikeda, Phys. Rev. B {\bf 83}, 224518 (2011). 
\bibitem{Gert} G. Eilenberger, Z. Physik {\bf 190}, 142-160 (1966). 
\bibitem{Tayama} The presence of a characteristic field which might be identified with this structural transition had been reported previously in relation to the magnetization measurement [Y. Namai, T. Tayama, T. Sakakibara, H. Shishido, Y. Haga, R. Settai, and Y. Onuki, in 59th Annual Meeting of the Physical Society of Japan (2004) (unpublished)]. 
\bibitem{RI76} R. Ikeda, Phys. Rev. B {\bf 76}, 134504 (2007). 
\bibitem{KMF} G. Koutroulakis, M. D. Stewart, Jr., V. F. Mitrovic, M. Horvatic, C. Berthier, G. Lapertot, and J. Flouquet, Phys. Rev. Lett. {\bf 104}, 087001 (2010). 
\bibitem{Volovik} G. E. Volovik, JETP Lett. {\bf 58}, 469 (1993). 
\bibitem{Mineev} V. P. Mineev, arXiv 1509.04915. 
\bibitem{Machida} K. M. Suzuki, M. Ichioka, and K. Machida, Phys. Rev. B {\bf 83}, 140503(R) (2011). 


\end{thebibliography}

\vspace{20mm}

\section{Appendix}

To evaluate $f^{(2, 2)}_{m, q_{\rm LO}}$, let us first expand the normal and anomalous Green's functions in powers of $q_{\rm LO}$. 
The O($q_{\rm LO}^2$) term of the normal Green's function 
\begin{equation}
{G}^{(\sigma)}_{\varepsilon_n, \, (0)}({\bm p}+{\bm Q}+ {\bm q}_{\rm LO} ,{\bm R}) 
=
\frac{-i\varepsilon_{n} - \varepsilon({\bm p}+{\bm Q}+{\bm q}_{\rm LO}) - \sigma I}
{\varepsilon_{n}^2 + \varepsilon^2({\bm p}+{\bm Q}+{\bm q}_{\rm LO}) 
+ |{\it \Delta}_{{\bm p}+{\bm Q}+ {\bm q}_{\rm LO}}({\bm R}) |^2 - I^2 - i \sigma 2 \varepsilon_n I}
,
\end{equation}
takes the form 
\begin{eqnarray}
{G}^{(\sigma)}_{\varepsilon_n, \, (0,2)}({\bm p}+{\bm Q}+ {\bm q}_{\rm LO}, {\bm R}) 
&=& 
- \frac{ i\varepsilon_{n} + \varepsilon({\bm p}+{\bm Q}) + \sigma I } 
{ ( a_2 - i \sigma b_1 )^3 }
\biggl[
2 \varepsilon ({\bm p}+{\bm Q})  {\bm q}_{\rm LO} \cdot {\bm v}_{{\bm p}+{\bm Q}} 
\nonumber\\
&+& {\bm q}_{\rm LO} \cdot (\nabla_{\bm p} |{\it \Delta}_{{\bm p}+{\bm Q}}({\bm R})|^2) 
\biggr]^2
+ \frac{i\varepsilon_{n} + \varepsilon({\bm p}+{\bm Q}) + \sigma I}
{ 2 ( a_2 - i \sigma b_1 )^2 }
\nonumber\\
&\times&
\biggl[ 2 
\varepsilon({\bm p}+{\bm Q})
({\bm q}_{\rm LO} \cdot \nabla_{\bm p})^2 \varepsilon({\bm p}+{\bm Q})
+ 2 ( {\bm q}_{\rm LO} \cdot {\bm v}_{{\bm p}+{\bm Q}} )^2 
\nonumber\\
&+& ({\bm q}_{\rm LO} \cdot \nabla_{\bm p})^2 |{\it \Delta}_{{\bm p}+{\bm Q}}({\bm R})|^2 
\biggr]
+ \frac{ {\bm q}_{\rm LO} \cdot {\bm v}_{{\bm p}+{\bm Q}} }
{ ( a_2 - i \sigma b_1 )^2 }
\nonumber\\
&\times&
\biggl[
2 \varepsilon({\bm p}+{\bm Q})  {\bm q}_{\rm LO} \cdot {\bm v}_{{\bm p}+{\bm Q}} 
+ {\rm q}_{\rm LO} (\nabla_{\bm p} |{\it \Delta}_{{\bm p}+{\bm Q}}({\bm R})|^2) 
\biggr]
\nonumber\\
&-& 
\frac{ ({\bm q}_{\rm LO} \cdot \nabla_{\bm p})^2 
\varepsilon({\bm p}+{\bm Q}) } 
{ 2 ( a_2 - i \sigma b_1 ) } .
\end{eqnarray}

Similarly, the O($q_{\rm LO}^2$) term of the anomalous Green's function 
\begin{equation}
{\overline F}^{(\sigma)}_{\varepsilon_n, \, (0)}({\bm p}+{\bm Q}+ {\bm q}_{\rm LO} ,{\bm R}) 
= 
\frac{-  \sigma {\it \Delta}^*_{{\bm p}+{\bm Q}+ {\bm q}_{\rm LO}}({\bm R})}
{\varepsilon_{n}^2 + \varepsilon^2({\bm p}+{\bm Q}+{\bm q}_{\rm LO}) 
+ |{\it \Delta}_{{\bm p}+{\bm Q}+ {\bm q}_{\rm LO}}({\bm R})|^2 - I^2 - i \sigma 2 \varepsilon_n I}
,
\end{equation}
is expressed in the form 
\begin{eqnarray}
{\overline F}^{(\sigma)}_{\varepsilon_n, \, (0,2)}({\bm p}+{\bm Q}+{\bm q}_{\rm LO} ,{\bm R}) 
&=& 
-  \frac{ \sigma {\it \Delta}^*_{{\bm p}+{\bm Q}}({\bm R}) } 
{ ( a_2 - i \sigma b_1 )^3 }
\biggl[2 
\varepsilon({\bm p}+{\bm Q})  {\bm q}_{\rm LO} \cdot {\bm v}_{{\bm p}+{\bm Q}} 
\nonumber\\
&+& {\bm q}_{\rm LO}\cdot(\nabla_{\bm p} |{\it \Delta}_{{\bm p}+{\bm Q}}({\bm R})|^2 ) \biggr]^2
+ \frac{\sigma {\it \Delta}^*_{{\bm p}+{\bm Q}}({\bm R})}
{ 2 ( a_2 - i \sigma b_1 )^2 }
\nonumber\\
&\times&
\biggl[ 2 
\varepsilon({\bm p}+{\bm Q})
({\bm q}_{\rm LO} \cdot \nabla_{\bm p})^2 \varepsilon({\bm p}+{\bm Q})
+ 2 ( {\bm q}_{\rm LO} \cdot {\bm v}_{{\bm p}+{\bm Q}} )^2 
\nonumber\\
&+& ({\bm q}_{\rm LO} \cdot \nabla_{\bm p})^2 |{\it \Delta}_{{\bm p}+{\bm Q}}({\bm R})|^2 
\biggr]
- \frac{ \sigma {\rm q}_{\rm LO} \cdot (\nabla_{\bm p} {\it \Delta}^{*}_{{\bm p}+{\bm Q}}({\bm R})) }
{ ( a_2 - i \sigma b_1 )^2 }
\nonumber\\
&\times&
\biggl[ 2 
\varepsilon({\bm p}+{\bm Q})  {\bm q}_{\rm LO} \cdot {\bm v}_{{\bm p}+{\bm Q}}
+ {\bm q}_{\rm LO} \cdot (\nabla_{\bm p} |{\it \Delta}_{{\bm p}+{\bm Q}} ({\bm R})|^2 ) 
\biggr]
\nonumber\\
&-& 
\frac{ \sigma ({\rm q}_{\rm LO} \cdot \nabla_{\bm p})^2  {\it \Delta}^*_{{\bm p}+{\bm Q}}({\bm R})} 
{ 2 ( a_2 - i \sigma b_1 ) }
, 
\end{eqnarray}
where 
\begin{eqnarray}
a_2 &=& \varepsilon_n^2+\varepsilon^2({\bm p}+{\bm Q})
+|{\it \Delta}_{{\bm p}+{\bm Q}}({\bm R})|^2-I^2, \nonumber \\
b_1 &=& 2 \varepsilon_n I. \nonumber \\
\end{eqnarray}

Using them, the first term of $f^{(2, 2)}_{m, q_{LO}}$ takes the form 
\begin{equation}
\biggl\langle
\frac{T}{2}
\sum^{}_{{\bm p},\sigma, \varepsilon_{n}} 
\sum^{}_{s_{1} = \pm 1} 
{\rm Tr}
\biggl[
{\hat G}^{(\sigma)}_{\varepsilon_n, \, (2)}({\bm p}+{\bm Q}_0+s_{1} {\bm q}, {\bm R})
{\hat G}^{(-\sigma)}_{\varepsilon_n, \, (0)}({\bm p}, {\bm R}) 
\biggr]
|m({\bm q}, {\bm R})|^{2} \biggr\rangle_{\rm sp} \equiv \langle C^{(2,0)}({\bm q}, {\bm R}) |m({\bm q}, {\bm R})|^2 \rangle_{\rm sp},
\end{equation}
where 
\begin{eqnarray}
C^{(2,0)}({\bm q}, {\bm R}) &=& 12T 
\sum^{}_{{\bm p}, n>0} 
\frac{ ({\bm v}_{{\bm p}+{\bm Q}} \cdot 
\nabla |{\it \Delta}_{{\bm p}+{\bm Q}}({\bm R})|^2)^2 }{c_{51}^2+d_{51}^2}
\biggl[ \Bigl\{
\varepsilon({\bm p}+{\bm Q})\varepsilon({\bm p})
[3\varepsilon_n^2 - \varepsilon^2({\bm p}+{\bm Q}) + 3|{\it \Delta}_{{\bm p}+{\bm Q}}({\bm R})|^2 -3I^2] 
 \nonumber \\
&-& [\varepsilon_n^2 - {\it \Delta}^*_{{\bm p}+{\bm Q}}({\bm R}) {\it \Delta}_{{\bm p}}({\bm R}) + I^2]
[\varepsilon_n^2 - 3\varepsilon^2({\bm p}+{\bm Q}) + |{\it \Delta}_{{\bm p}+{\bm Q}}({\bm R})|^2 - I^2]
\Bigr\}  c_{51}
\nonumber \\
&+&[\varepsilon_n^2 - 3\varepsilon({\bm p}+{\bm Q}) \varepsilon({\bm p}) 
- {\it \Delta}^*_{{\bm p}+{\bm Q}}({\bm R}) {\it \Delta}_{{\bm p}}({\bm R}) +I^2 ] b_1 d_{51}
\biggr]
\nonumber \\
&+&
4T 
\sum^{}_{{\bm p}, n>0} 
\frac{ ({\bm v}_{{\bm p}+{\bm Q}} \cdot 
\nabla |{\it \Delta}_{{\bm p}+{\bm Q}}({\bm R})|^2)^2 }{c_{41}^2+d_{41}^2}
[\varepsilon_n^2 - 4\varepsilon({\bm p}+{\bm Q})\varepsilon({\bm p}) 
- {\it \Delta}^*_{{\bm p}+{\bm Q}}({\bm R}) {\it \Delta}_{{\bm p}}({\bm R}) + I^2
]c_{41}
\nonumber \\
&-&
8T 
\sum^{}_{{\bm p}, n>0} 
\frac{ ({\bm v}_{{\bm p}+{\bm Q}} \cdot 
\nabla |{\it \Delta}_{{\bm p}+{\bm Q}}({\bm R})|^2)
( {\it \Delta}_{{\bm p}} {\bm v}_{{\bm p}+{\bm Q}} \cdot 
\nabla {\it \Delta}^*_{{\bm p}+{\bm Q}}({\bm R}) )
 }{c_{41}^2+d_{41}^2}
 \nonumber \\
&\times&
\biggl[ 
[\varepsilon_n^2 - 2\varepsilon^2({\bm p}+{\bm Q}) + |{\it \Delta}_{{\bm p}+{\bm Q}}({\bm R})|^2 - I^2] c_{41}
- b_1 d_{41}
\biggr]
\nonumber \\
&-&
4T 
\sum^{}_{{\bm p}, n>0} 
\frac{ ({\bm v}_{{\bm p}+{\bm Q}} \cdot \nabla )^2
|{\it \Delta}_{{\bm p}+{\bm Q}}({\bm R})|^2 }{c_{41}^2+d_{41}^2}
\biggl[  \Bigl\{ 
\varepsilon({\bm p}+{\bm Q})\varepsilon({\bm p})
[3\varepsilon_n^2 + \varepsilon^2({\bm p}+{\bm Q}) + 3|{\it \Delta}_{{\bm p}+{\bm Q}}({\bm R})|^2 -3I^2] 
 \nonumber \\
&-& [\varepsilon_n^2 - {\it \Delta}^*_{{\bm p}+{\bm Q}}({\bm R}) {\it \Delta}_{{\bm p}}({\bm R}) + I^2]
[\varepsilon_n^2 - 3\varepsilon^2({\bm p}+{\bm Q}) + |\Delta_{{\bm p}+{\bm Q}}({\bm R})|^2 - I^2]
\Bigr\} c_{41}
\nonumber \\
&+&
[\varepsilon_n^2 - 3\varepsilon({\bm p}+{\bm Q}) \varepsilon({\bm p}) 
- {\it \Delta}^*_{{\bm p}+{\bm Q}}({\bm R}) {\it \Delta}_{{\bm p}}({\bm R}) -I^2 ] b_1 d_{41}
\biggr]
\nonumber \\
&-&
4T 
\sum^{}_{{\bm p}, n>0} 
\frac{ |{\bm v}_{{\bm p}+{\bm Q}} \cdot \nabla {\it \Delta}_{{\bm p}+{\bm Q}}({\bm R})|^2 }
{c_{31}^2+d_{31}^2}
[\varepsilon_n^2 - \varepsilon({\bm p}+{\bm Q})\varepsilon({\bm p}) 
- {\it \Delta}^*_{{\bm p}+{\bm Q}}({\bm R}) {\it \Delta}_{{\bm p}}({\bm R}) + I^2
]c_{31}
\nonumber \\
&+&
8T 
\sum^{}_{{\bm p}, n>0} 
\frac{{\it \Delta}_{{\bm p}+{\bm Q}} ({\bm R})
( {\bm v}_{{\bm p}+{\bm Q}}({\bm R}) \cdot \nabla )^2 {\it \Delta}^*_{{\bm p}+{\bm Q}}({\bm R}) }
{c_{31}^2+d_{31}^2}
\varepsilon({\bm p}+{\bm Q})\varepsilon({\bm p}) c_{31}
\nonumber \\
&+&
4T 
\sum^{}_{{\bm p}, n>0} 
\frac{{\it \Delta}_{{\bm p}} ({\bm R})
( {\bm v}_{{\bm p}+{\bm Q}} \cdot \nabla )^2 {\it \Delta}^*_{{\bm p}+{\bm Q}}({\bm R}) }
{c_{31}^2+d_{31}^2}
\biggl[ 
[\varepsilon_n^2 - \varepsilon^2({\bm p}+{\bm Q}) + |{\it \Delta}_{{\bm p}+{\bm Q}}({\bm R})|^2 - I^2] c_{31}
- b_1 d_{31}
\biggr],
\end{eqnarray}

\vspace{22in}

The corresponding second term is 
\begin{equation}
\biggl\langle
\frac{T}{2}
\sum^{}_{{\bm p},\sigma, \varepsilon_{n}} 
\sum^{}_{s_{1} = \pm 1} 
{\rm Tr}
\biggl[
{\hat G}^{(\sigma)}_{\varepsilon_n, \, (1)}({\bm p}+{\bm Q}_0+s_{1} {\bm q}, {\bm R})
{\hat G}^{(-\sigma)}_{\varepsilon_n, \, (1)}({\bm p}, {\bm R})
\biggr]
|m({\bm q}, {\bm R})|^{2} \biggr\rangle_{\rm sp} \equiv \langle C^{(1,1)}({\bm q}, {\bm R}) |m({\bm q}, {\bm R})|^2 \rangle_{\rm sp},
\end{equation}
where 
\begin{eqnarray}
C^{(1,1)}({\bm q}, {\bm R}) &=& - 4T 
\sum^{}_{{\bm p}, n>0} 
\frac{ ({\bm v}_{{\bm p}+{\bm Q}} \cdot 
\nabla |{\it \Delta}_{{\bm p}+{\bm Q}}({\bm R})|^2) ({\bm v}_{{\bm p}} \cdot 
\nabla |{\it \Delta}_{{\bm p}}({\bm R})|^2) }{c_{33}^2+d_{33}^2}
\biggl[ \Bigl\{
[\varepsilon_n^2 - \varepsilon^2({\bm p}+{\bm Q}) + |{\it \Delta}_{{\bm p}+{\bm Q}}({\bm R})|^2 - I^2 ]
\nonumber \\
&\times& [\varepsilon_n^2 - \varepsilon^2({\bm p}) + |{\it \Delta}_{{\bm p}}({\bm R})|^2 - I^2] 
-4( \varepsilon_n^2 - {\it \Delta}^*_{{\bm p}+{\bm Q}}({\bm R}) {\it \Delta}_{{\bm p}}({\bm R}) + I^2 )
\varepsilon({\bm p}+{\bm Q}) \varepsilon({\bm p}) + 4 \varepsilon_n^2 I^2 \Bigr\} c_{33}
\nonumber \\
&-&
[\varepsilon^2({\bm p}+{\bm Q}) - \varepsilon^2({\bm p}) 
- |{\it \Delta}_{{\bm p}+{\bm Q}}({\bm R})|^2 +  |{\it \Delta}_{{\bm p}}({\bm R})|^2  ] b_1 d_{51}
\biggr]
\nonumber \\
&+&
2T 
\sum^{}_{{\bm p}, n>0} 
\frac{ ({\bm v}_{{\bm p}+{\bm Q}} \cdot 
\nabla |{\it \Delta}_{{\bm p}+{\bm Q}}({\bm R})|^2) ({\bm v}_{{\bm p}} \cdot 
\nabla |{\it \Delta}_{{\bm p}}({\bm R})|^2) }{c_{32}^2+d_{32}^2}
\biggl[ 
[\varepsilon_n^2 - \varepsilon^2({\bm p}+{\bm Q}) + |{\it \Delta}_{{\bm p}+{\bm Q}}({\bm R})|^2 - I^2 ] c_{32}
- b_1 d_{32}
\biggr]
\nonumber \\
&+&
8T 
\sum^{}_{{\bm p}, n>0} 
\frac{ ({\bm v}_{{\bm p}+{\bm Q}} \cdot \nabla |{\it \Delta}_{{\bm p}+{\bm Q}}({\bm R})|^2) 
 ({\it \Delta}^*_{{\bm p}+{\bm Q}}({\bm R}) {\bm v}_{\bm p} \cdot  \nabla {\it \Delta}_{\bm p}({\bm R})) }
{c_{32}^2+d_{32}^2}
\varepsilon({\bm p}+{\bm Q}) \varepsilon({\bm p}) c_{32}
\nonumber \\
&+&
2T 
\sum^{}_{{\bm p}, n>0} 
\frac{ ({\bm v}_{{\bm p}+{\bm Q}} \cdot 
\nabla |{\it \Delta}_{{\bm p}+{\bm Q}}({\bm R})|^2) ({\bm v}_{\bm p} \cdot 
\nabla |{\it \Delta}_{{\bm p}}({\bm R})|^2) }{c_{23}^2+d_{23}^2}
\biggl[ 
[\varepsilon_n^2 - \varepsilon^2({\bm p}) + |{\it \Delta}_{{\bm p}}({\bm R})|^2 - I^2 ] c_{23}
+ b_1 d_{23}
\biggr]
\nonumber \\
&+&
8T 
\sum^{}_{{\bm p}, n>0} 
\frac{ ({\bm v}_{{\bm p}} \cdot \nabla |{\it \Delta}_{{\bm p}}({\bm R})|^2) 
 ({\it \Delta}^*_{{\bm p}}({\bm R}) {\bm v}_{{\bm p}+{\bm Q}}
\cdot  \nabla {\it \Delta}_{{\bm p}+{\bm Q}}({\bm R})) }
{c_{23}^2+d_{23}^2}
\varepsilon({\bm p}+{\bm Q}) \varepsilon({\bm p}) c_{23}
\nonumber \\
&-&
T 
\sum^{}_{{\bm p}, n>0} 
\frac{ ({\bm v}_{{\bm p}+{\bm Q}} \cdot 
\nabla |{\it \Delta}_{{\bm p}+{\bm Q}}({\bm R})|^2) ({\bm v}_{{\bm p}} \cdot 
\nabla |{\it \Delta}_{{\bm p}}({\bm R})|^2) }
{c_{22}^2+d_{22}^2}
 c_{22}
\nonumber \\
&-&
4T 
\sum^{}_{{\bm p}, n>0} 
\frac{ ({\bm v}_{{\bm p}+{\bm Q}} \cdot 
\nabla {\it \Delta}^*_{{\bm p}+{\bm Q}}({\bm R}) ) ({\bm v}_{{\bm p}} \cdot 
\nabla {\it \Delta}_{{\bm p}}({\bm R})) }
{c_{22}^2+d_{22}^2}
[\varepsilon_n^2 + \varepsilon({\bm p}+{\bm Q}) \varepsilon({\bm p}) + I^2 ] c_{22}.
\end{eqnarray}

\vspace{22in}

Further, its third term becomes 
\begin{equation}
\biggl\langle
\frac{T}{2}
\sum^{}_{{\bm p},\sigma, \varepsilon_{n}} 
\sum^{}_{s_{1} = \pm 1} 
{\rm Tr}
\biggl[
{\hat G}^{(\sigma)}_{\varepsilon_n, \, (0, 0)}({\bm p}+{\bm Q}_0+s_{1} {\bm q}, {\bm R})
{\hat G}^{(-\sigma)}_{\varepsilon_n, \, (2)}({\bm p}, {\bm R})
\biggr]
|m({\bm q}, {\bm R})|^{2} \biggr\rangle_{\rm sp} \equiv \langle C^{(0,2)}({\bm q}, {\bm R}) |m({\bm q}, {\bm R})|^2 \rangle_{\rm sp}, 
\end{equation}
where 
\begin{eqnarray}
C^{(0,2)}({\bm q}, {\bm R}) &=& 12T 
\sum^{}_{{\bm p}, n>0} 
\frac{ ({\bm v}_{{\bm p}} \cdot 
\nabla |{\it \Delta}_{{\bm p}}({\bm R})|^2)^2 }{c_{15}^2+d_{15}^2}
\biggl[  \Bigl\{
\varepsilon({\bm p}+{\bm Q})\varepsilon({\bm p})
[3\varepsilon_n^2 - \varepsilon^2({\bm p}) + 3|{\it \Delta}_{{\bm p}}({\bm R})|^2 -3I^2] 
 \nonumber \\
&-& [\varepsilon_n^2 - {\it \Delta}^*_{{\bm p}+{\bm Q}}({\bm R}) {\it \Delta}_{{\bm p}}({\bm R}) + I^2]
[\varepsilon_n^2 - 3\varepsilon^2({\bm p}) + |{\it \Delta}_{{\bm p}}({\bm R})|^2 - I^2] \Bigr\} c_{15}
\nonumber \\
&-&[\varepsilon_n^2 - 3\varepsilon({\bm p}+{\bm Q}) \varepsilon({\bm p}) 
- {\it \Delta}^*_{{\bm p}+{\bm Q}}({\bm R}) {\it \Delta}_{{\bm p}}({\bm R}) + I^2 ] b_1 d_{15}
\biggr]
\nonumber \\
&+&
4T 
\sum^{}_{{\bm p}, n>0} 
\frac{ ({\bm v}_{{\bm p}} \cdot 
\nabla |{\it \Delta}_{{\bm p}}({\bm R})|^2)^2 }{c_{14}^2+d_{14}^2}
[\varepsilon_n^2 - 4\varepsilon({\bm p}+{\bm Q})\varepsilon({\bm p}) 
- {\it \Delta}^*_{{\bm p}+{\bm Q}}({\bm R}) {\it \Delta}_{{\bm p}}({\bm R}) + I^2
]c_{14}
\nonumber \\
&-&
8T 
\sum^{}_{{\bm p}, n>0} 
\frac{ ({\bm v}_{{\bm p}} \cdot 
\nabla |{\it \Delta}_{{\bm p}}({\bm R})|^2)
( {\it \Delta}^*_{{\bm p}+{\bm Q}}({\bm R}) {\bm v}_{{\bm p}} \cdot 
\nabla {\it \Delta}_{{\bm p}}({\bm R}) )
 }{c_{14}^2+d_{14}^2} 
\biggl[ 
[\varepsilon_n^2 - 2\varepsilon^2({\bm p}) + |{\it \Delta}_{{\bm p}}({\bm R})|^2 - I^2] c_{14}
+ b_1 d_{14}
\biggr]
\nonumber \\
&-&
4T 
\sum^{}_{{\bm p}, n>0} 
\frac{ ({\bm v}_{{\bm p}} \cdot \nabla )^2
|{\it \Delta}_{{\bm p}}({\bm R})|^2 }{c_{14}^2+d_{14}^2}
\biggl[ \Bigl\{
\varepsilon({\bm p}+{\bm Q})\varepsilon({\bm p})
[3\varepsilon_n^2 - \varepsilon^2({\bm p}) + 3|{\it \Delta}_{{\bm p}}({\bm R})|^2 -3I^2] 
 \nonumber \\
&-& [\varepsilon_n^2 - {\it \Delta}^*_{{\bm p}+{\bm Q}}({\bm R}) {\it \Delta}_{{\bm p}}({\bm R}) + I^2]
[\varepsilon_n^2 - 3\varepsilon^2({\bm p}) + |{\it \Delta}_{{\bm p}}({\bm R})|^2 - I^2] \Bigr\} c_{14}
\nonumber \\
&-&
[\varepsilon_n^2 - 3\varepsilon({\bm p}+{\bm Q}) \varepsilon({\bm p}) 
- {\it \Delta}^*_{{\bm p}+{\bm Q}}({\bm R}) {\it \Delta}_{{\bm p}}({\bm R}) -I^2 ] b_1 d_{14}
\biggr]
\nonumber \\
&-&
4T 
\sum^{}_{{\bm p}, n>0} 
\frac{ |{\bm v}_{{\bm p}} \cdot \nabla {\it \Delta}_{{\bm p}}({\bm R})|^2 }
{c_{13}^2+d_{13}^2}
[\varepsilon_n^2 - \varepsilon({\bm p}+{\bm Q})\varepsilon({\bm p}) 
- {\it \Delta}^*_{{\bm p}+{\bm Q}}({\bm R}) {\it \Delta}_{{\bm p}}({\bm R}) + I^2
]c_{13}
\nonumber \\
&+&
8T 
\sum^{}_{{\bm p}, n>0} 
\frac{ {\it \Delta}_{{\bm p}}({\bm R}) 
( {\bm v}_{{\bm p}} \cdot \nabla  )^2 {\it \Delta}^*_{{\bm p}}({\bm R}) }
{c_{13}^2+d_{13}^2}
\varepsilon({\bm p}+{\bm Q})\varepsilon({\bm p}) c_{13}
\nonumber \\
&+&
4T 
\sum^{}_{{\bm p}, n>0} 
\frac{{\it \Delta}^*_{{\bm p}+{\bm Q}}({\bm R}) 
( {\bm v}_{{\bm p}} \cdot \nabla )^2 {\it \Delta}_{{\bm p}}({\bm R}) }
{c_{31}^2+d_{31}^2}
\biggl[ 
[\varepsilon_n^2 - \varepsilon^2({\bm p}) + |{\it \Delta}_{{\bm p}}({\bm R})|^2 - I^2] c_{13}
+ b_1 d_{13}
\biggr]
.
\end{eqnarray}

Finally, the fourth term is 
\begin{equation}
\biggl\langle
\frac{T}{2}
\sum^{}_{{\bm p},\sigma, \varepsilon_{n}} 
\sum^{}_{s_{1} = \pm 1} 
{\rm Tr}
\biggl[
{\hat G}^{(\sigma)}_{\varepsilon_n, \, (0, 2)}({\bm p}+{\bm Q}_0+s_{1} {\bm q}, {\bm R})
{\hat G}^{(-\sigma)}_{\varepsilon_n, \, (0)}({\bm p}, {\bm R}) 
\biggr]
|m({\bm q}, {\bm R})|^{2} \biggr\rangle_{\rm sp} \equiv \langle C^{(2)}({\bm q}, {\bm R}) |m({\bm q}, {\bm R})|^{2} \rangle_{\rm sp}, 
\end{equation}
where 
\begin{eqnarray}
C^{(2)}({\bm q}, {\bm R}) &=& 
- 4 T \sum^{}_{{\bm p}, n>0} 
\frac{ [ 
2 {\bm q}_{\rm LO} \cdot {\bm v}_{{\bm p}+{\bm Q}} 
\varepsilon({\bm p}+{\bm Q})
+ {\bm q}_{\rm LO} \cdot (\nabla_{\bm p} |{\it \Delta}_{{\bm p}+{\bm Q}}({\bm R})|^2 ) ]^2 }
{c_{31}^2+d_{31}^2}
\nonumber \\
&\times&
[ \varepsilon_n^2 - \varepsilon({\bm p}+{\bm Q}) \varepsilon({\bm p}) 
- {\it \Delta}^*_{{\bm p}+{\bm Q}}({\bm R}) {\it \Delta}_{{\bm p}}({\bm R}) + I^2] c_{31}
\nonumber \\
&+& 2T
\sum^{}_{{\bm p}, n>0} 
\frac{ 2 
\varepsilon({\bm p}+{\bm Q})({\bm q}_{\rm LO} \cdot \nabla_{\bm p})^2 \varepsilon({\bm p}+{\bm Q}) 
+  ({\bm q}_{\rm LO} \cdot \nabla_{\bm p})^2 |{\it \Delta}_{{\bm p}+{\bm Q}}({\bm R})|^2
}{c_{21}^2+d_{21}^2}
\nonumber \\
&\times&
[ \varepsilon_n^2 - \varepsilon({\bm p}+{\bm Q}) \varepsilon({\bm p}) 
- {\it \Delta}^*_{{\bm p}+{\bm Q}}({\bm R}) {\it \Delta}_{{\bm p}}({\bm R}) + I^2] c_{21}
\nonumber \\
&+& 4T
\sum^{}_{{\bm p}, n>0} 
\frac{ 
[{\bm q}_{\rm LO} \cdot {\bm v}_{{\bm p}+{\bm Q}} ]^2 }{c_{21}^2+d_{21}^2}
[ \varepsilon_n^2 - 3\varepsilon({\bm p}+{\bm Q}) \varepsilon({\bm p}) 
- {\it \Delta}^*_{{\bm p}+{\bm Q}}({\bm R}) {\it \Delta}_{{\bm p}}({\bm R}) + I^2] c_{21}
\nonumber \\
&+& 2T
\sum^{}_{{\bm p}, n>0} 
\frac{ 
\varepsilon({\bm p}) ({\bm q}_{\rm LO} \cdot \nabla_{\bm p})^2  \varepsilon({\bm p}+{\bm Q}) + {\it \Delta}_{\bm p}({\bm R}) ({\bm q}_{\rm LO} \cdot \nabla_{\bm p})^2 {\it \Delta}^*_{{\bm p}+{\bm Q}}({\bm R})}{c_{11}^2+d_{11}^2}
c_{11}
\nonumber \\
&-& 4T
\sum^{}_{{\bm p}, n>0} \frac{c_{21}}{c_{21}^2+d_{21}^2} [({\bm q}_{\rm LO} \cdot {\bm v}_{{\bm p}+{\bm Q}}) [{\bm q}_{\rm LO} \cdot (\nabla_{\bm p}|{\it \Delta}_{{\bm p}+{\bm Q}}({\bm R})|^2)] \varepsilon({\bm p}) \nonumber \\
&+& 2 {\it \Delta}_{\bm p}({\bm R}) [ {\bm q}_{\rm LO} \cdot 
\nabla_{\bm p} {\it \Delta}^*_{{\bm p}+{\bm Q}}({\bm R}) ] ({\bm q}_{\rm LO} \cdot {\bm v}_{{\bm p}+{\bm Q}}) \varepsilon({\bm p}+{\bm Q}) \nonumber \\
&+& {\it \Delta}_{\bm p}({\bm R}) [{\bm q}_{\rm LO} \cdot (\nabla_{\bm p}|{\it \Delta}_{{\bm p}+{\bm Q}}({\bm R})|^2)] [{\bm q}_{\rm LO} \cdot (\nabla_{\bm p} {\it \Delta}^*_{{\bm p}+{\bm Q}}({\bm R})) ] ].
\end{eqnarray}

The coefficients appeared in the above expressions are given by 
\begin{eqnarray}
a_1 &=& \varepsilon_n^2+\varepsilon^2({\bm p})+|{\it \Delta}_{\bm p}({\bm R})|^2-I^2, \nonumber \\
a_2 &=& \varepsilon_n^2+\varepsilon^2({\bm p}+{\bm Q})
+|{\it \Delta}_{{\bm p}+{\bm Q}}({\bm R})|^2-I^2, \nonumber \\
b_1 &=& 2 \varepsilon_n I, \nonumber \\
c_{11} &=& a_{2}a_{1} + b_{1}^{2}, \nonumber \\ 
d_{11} &=& ( a_{2} - a_{1} )b_{1} , \nonumber \\ 
c_{21} &=& (a_2^2 - b_1^2)a_1 + (2a_2 b_{1}) b_{1}, \nonumber \\ 
d_{21} &=& (a_2^2 - b_1^2)b_1 - (2a_2 b_{1}) a_{1}, \nonumber \\ 
c_{31} &=& (a_2^3 - 3a_2 b_1^2)a_1 + (3a_2^2 b_1 - b_1^3)b_1, \nonumber \\ 
d_{31} &=& (a_2^3 - 3a_2 b_1^2)b_1 - (3a_2^2 b_1 - b_1^3)a_1, \nonumber \\
c_{41} &=& (a_2^4 - 6a_2^2 b_1^2 + b_1^4)a_1 + (4a_2^3 b_1 - 4a_2 b_1^3)b_1, \nonumber \\ 
d_{41} &=& (a_2^4 - 6a_2^2 b_1^2 + b_1^4)b_1 - (4a_2^3 b_1 - 4a_2 b_1^3)a_1, \nonumber \\
c_{51} &=& (a_2^5 - 10a_2^3 b_1^2 + 5a_2 b_1^4)a_1 + (5a_2^4 b_1 - 10a_2^2 b_1^3 + b_1^5)b_1, \nonumber \\ 
d_{51} &=& (a_2^5 - 10a_2^3 b_1^2 + 5a_2 b_1^4)b_1 - (5a_2^4 b_1 - 10a_2^2 b_1^3 + b_1^5)a_1, \nonumber \\
c_{12} &=& a_2(a_1^2 - b_1^2) + b_{1}(2a_1 b_{1}), \nonumber \\ 
d_{12} &=& b_1(a_1^2 - b_1^2) - a_{2}(2a_1 b_{1})  , \nonumber \\ 
c_{13} &=& a_2(a_1^3 - 3a_1 b_1^2) + b_1(3a_1^2 b_1 - b_1^3), \nonumber \\ 
d_{13} &=& b_1(a_1^3 - 3a_1 b_1^2) - a_2(3a_1^2 b_1 - b_1^3), \nonumber \\
c_{14} &=& a_2(a_1^4 - 6a_1^2 b_1^2 + b_1^4) + b_1(4a_1^3 b_1 - 4a_1 b_1^3), \nonumber \\ 
d_{14} &=& b_1(a_1^4 - 6a_1^2 b_1^2 + b_1^4) - a_2(4a_1^3 b_1 - 4a_1 b_1^3), \nonumber \\
c_{15} &=& a_2(a_1^5 - 10a_1^3 b_1^2 + 5a_1 b_1^4) + b_1(5a_1^4 b_1 - 10a_1^2 b_1^3 + b_1^5), \nonumber \\ 
d_{15} &=& b_1(a_1^5 - 10a_1^3 b_1^2 + 5a_1 b_1^4) - a_2(5a_1^4 b_1 - 10a_1^2 b_1^3 + b_1^5), \nonumber \\
c_{22} &=& (a_2^2 - b_1^2)(a_1^2 - b_1^2) + (2a_2 b_1)(2a_1 b_1), \nonumber \\ 
d_{22} &=& (a_2^2 - b_1^2)(2a_1 b_1) - (2a_2 b_1)(a_1^2 - b_1^2)  , \nonumber \\ 
c_{32} &=& (a_2^3 - 3a_2 b_1^2)(a_1^2 - b_1^2) + (3a_2^2 b_1 - b_1^3)(2a_1 b_1), \nonumber \\ 
d_{32} &=& (a_2^3 - 3a_2 b_1^2)(2a_1 b_1) - (3a_2^2 b_1 - b_1^3)(a_1^2 - b_1^2) , \nonumber \\ 
c_{23} &=& (2a_2 b_1)(a_1^3 - 3a_1 b_1^2) + (2a_2 b_1)(3a_2^2 b_1 - b_1^3), \nonumber \\ 
d_{23} &=& (2a_2 b_1)(3a_2^2 b_1 - b_1^3) - (2a_2 b_1)(a_1^3 - 3a_1 b_1^2) , \nonumber \\ 
c_{33} &=& (a_2^3 - 3a_2 b_1^2)(a_1^3 - 3a_1 b_1^2) + (3a_2^2 b_1 - b_1^3)(3a_1^2 b_1 - b_1^3), \nonumber \\ 
d_{33} &=& (a_2^3 - 3a_2 b_1^2)(3a_1^2 b_1 - b_1^3) - (3a_2^2 b_1 - b_1^3)(a_1^3 - 3a_1 b_1^2).  
\end{eqnarray}

\end{document}